\documentclass[a4paper,fleqn,usenatbib]{mnras}

\usepackage{newtxtext,newtxmath}
\usepackage[T1]{fontenc}
\usepackage{ae,aecompl}

\usepackage{graphicx}	
\usepackage{amsmath}	
\usepackage{amssymb}	

\usepackage{xcolor}
\definecolor{brownRed}{rgb}{0.65, 0.16, 0.16}
\definecolor{EngGreen}{rgb}{0.0, 0.5, 0.0}
\definecolor{azure}{rgb}{0.0, 0.5, 1.0}
\definecolor{darkblue}{rgb}{0.0, 0.0, 0.55}
\definecolor{Cred}{rgb}{0.7, 0.11, 0.01}
\definecolor{Dgreen}{rgb}{0.0, 0.5, 0.0}

\title[AGN feedback and SMBHB shrinking]{The effect of AGN feedback
  on the migration timescale of supermassive black holes binaries}
\author[del Valle et al.]{
  Luciano del Valle,$^{1}$\thanks{E-mail: delvalle@iap.fr }
and Marta Volonteri$^{1}$
\\
$^{1}$ Institut d'Astrophysique de Paris, Sorbonne Universit\'e, UPMC Univ
 CNRS, UMR 7095, 98 bis bd Arago, 75014 Paris, France
}

\date{Accepted XXX. Received YYY; in original form ZZZ}

\pubyear{2017}

\begin{document}
\maketitle

\begin{abstract}

  The gravitational interaction at parsec to sub-parsec scales between a circumbinary
  gas disc and a super massive black hole binary (SMBHB) is a promising mechanism to
  drive the migration of SMBHBs toward coalescence. The typical dynamical evolution can
  be separated in two main regimes:
  I) Slow migration ($T_{\rm mig}$ $\sim$ $10^{3-4}\times T_{\rm orb}$), where
  viscous torques are not efficient enough to redistribute the extracted angular
  momentum from the binary, leading to the formation of a low density cavity
  around the binary. II) Fast migration  ($T_{\rm mig}$ $\sim$ $10^{1-2}
  \times T_{\rm orb}$), in which the redistribution of angular momentum is
  efficient  and no low density cavity is formed in the circumbinary disc.
  Using {\it N}-Body/SPH simulations we study the effect of AGN feedback in
  this phase of a SMBHB evolution. We implement an
  AGN feedback model in the SPH code Gadget-3 that includes momentum
  feedback from winds, X-ray heating/radial-momentum and Eddington force.
  Using this implementation we run a set of simulations of SMBHB+disc in the
  two main shrinking regimes. From these simulations we conclude that the
  effect of the AGN mechanical feedback is negligible for SMBHBs in the slowly
  shrinking regime. However, in the fast shrinking regime the AGN wind excavate a
  ``feedback cavity'' leaving the SMBHB almost naked, thus stalling the orbital decay of the binary.
   
\end{abstract}

\begin{keywords}
black hole binary -- AGN -- gravitational waves
\end{keywords}



\section{Introduction}

Central SMBHs are found in practically every galaxy
with a significant bulge \citep{RICH1998,MAG1998,GULT2009}.
Within the currently accepted evolutionary model of the Universe 
the merger between galaxies is a common event 
\citep{WhiteFrenk1991}.

In mergers of galaxies with similar mass, if each galaxy involved
in the merger hosts a SMBH, both SMBHs will sink by dynamical 
friction to the innermost region of the core of the merger 
remnant \citep[e.g.][and references therein]{Colpi2014,Volonteri2016}.
When the mass enclosed by the orbit of these two SMBHs
is smaller than the sum of their masses, they will
form a SMBH binary \citep[][and reference therein]{Pfister2017}.
Understanding the further evolution of these SMBH binaries (SMBHBs)
is crucial because if two SMBHs are able to reach separations of 
$a_{\rm GW} \sim 10^{-3} (M_{\rm SMBHs}/10^6 M_{\odot})$ pc, then
the binary becomes a strong emitter of gravitational waves,
which leads to its coalescence \citep{PetersM1963,Peters1964}.
Gravitational waves from SMBHs are a prime source in the frequency
range of the future Laser Interferometer Space Antenna
\citep[LISA,][]{LISA2017}.

If the binary is formed in a gas-rich environment it is typically assumed
that its evolution leads to the formation of a circumbinary disc around the SMBHB.
These discs have been produced in the final stage of simulations of SMBHB formation
inside massive nuclear discs \citep{delValle2015}.
Recently, in the context of the accretion of clumpy
cold gas on to SMBH binaries, it has been shown that
a continuous supply of incoherent clouds produces an
intermittent formation and disruption of circumbinary
structures \citep{MaureiraFredes2018,Goicovic2018}.

The dynamical evolution of a SMBHB embedded in a circumbinary disc
is driven by gravitational torques produced by the disc on to the binary.
A variety of studies on the interaction between binaries
and discs in a broad range of contexts show that the angular momentum
extracted from the binary is redistributed in the disc
by means of viscous diffusion \citep[e.g.][]{GT1980, AR2005, Shi2012}.
If this redistribution is inefficient then the binary will
excavate a tidal gap/cavity of low density in the disc, resulting
in a drop of the intensity of the gravitational torque exerted
on to the binary and therefore the binary evolves slowly
($(da_{\rm bin}/dt)\,(1/a_{\rm bin})\sim 1/(10^3\,t_{\rm orb})$).
Instead, if the viscous diffusion in the disc redistributes
the extracted angular momentum rapidly enough, then no tidal gap/cavity
formation occurs and the binary evolves on a shorter timescale
($(da_{\rm bin}/dt)\,(1/a_{\rm bin})\sim 1/t_{\rm orb}$) \citep{LP1986, Escala2005, Cuadra2009}.
Motivated by this relation between the formation of a
tidal gap/cavity in the disc and the shrinking rate of the binary,
\cite{delValle2012,delValle2014} derived a gap/cavity opening
criterion for comparable mass binaries ($q\sim1$) comparing the
timescales for opening the tidal gap/cavity, driven by the
gravitational torque of the binary, with the timescale for closing
the tidal gap/cavity, driven by the viscous diffusion on the disc.

  In this work we use \cite{delValle2012,delValle2014} criterion to select binary-disc systems
in the two migration regimes and we study the effect of SMBH accretion/feedback
in both regimes. The effect of AGN feedback on the dynamical evolution of SMBHs has been
studied in the context of the inspiral of SMBH pairs in a star forming
massive nuclear disc \citep{SouzaLima} and in the context
of gravitationally recoiled SMBHs \citep{Sijacki}.
Both studies have found that AGN thermal feedback can disturb the
wake produced by the SMBHs on the background gas, decreasing significantly
the effect of dynamical friction on to the SMBHs.
These works are consistent with the work of \cite{Park} in which they studied
the effect of AGN radiation on the dynamical friction experienced by an accreting
SMBH that travels, with constant velocity, in a homogeneous distribution of gas.
They found that if the HII region around the SMBH is larger than the wake
produced by the black hole (that is comparable to the gravitational influence
sphere of the black hole) then dynamical friction is totally suppressed.
All of these studies suggest that AGN feedback can be a crucial
aspect of the dynamical evolution of SMBHs. We explore further this idea by
studying the effect of AGN feedback on the evolution of SMBHs
binaries at small separation, in the regime where dynamical friction is not 
the dominant process driving the dynamics. 

The structure of this paper is as follows: In section \S 2 we
discuss how we constrain and select the initial
conditions, in section \S 3 we present the numerical method and
parameters that we use in the simulations, in section \S 4 we present the
results and finally in section \S 5 we outline the main results
and conclusions of this study.

\section{Constraints on Circumbinary Discs Parameters}
\label{ICS_CONTRAINTS}

\begin{figure}
	\includegraphics[width=\columnwidth]{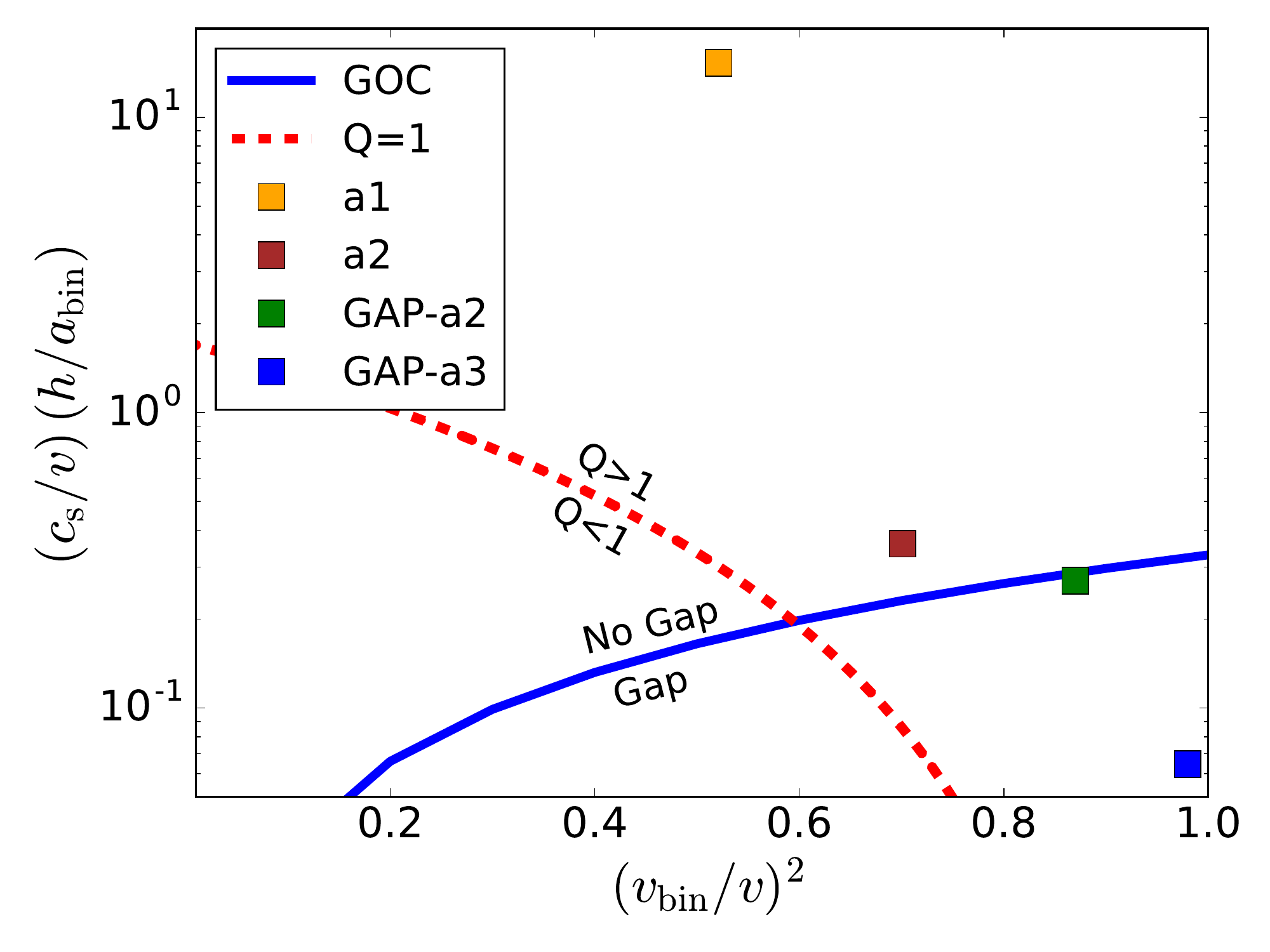}
        \caption{Representation of the gap/cavity opening criterion
          in the space of parameters
          $\left(\,(v_{\rm bin}/v)^2\,,\,(c_{\rm s}/v)(h/a_{\rm bin})\right)$
          (blue thick line).
          Above/below the blue line the space of parameter corresponds
          to binary-disc systems without/with a cavity. We also plot
          the $Q$ Toomre stability parameter for two values of the
          parameter $\beta=(c_{\rm s}/v)(a_{\rm bin}/h)$ ($\beta\approx 1$ for
          rotational supported systems).
          The square represents the parameters of the initial conditions
          of the simulations that we use in this study.
        }
    \label{GOC}
\end{figure}

We use the tidal gap/cavity opening criterion of \citet{delValle2014}
to select the initial conditions for the simulations of this
study. This criterion can be expressed as
\begin{eqnarray}
  \frac{\Delta t_{\rm open}}{\Delta t_{\rm close}}&=&\frac{1}{0.33}\,
  \left(\frac{v}{v_{\rm bin}}\right)^2\,
  \left(\frac{c_{\rm s}}{v}\right) \,\left(\frac{h}{a_{\rm bin}}\right)
  \;\le\; 1 \,,
\label{GapOpeningA}
\end{eqnarray}
where $c_{\rm s}$ is the sound speed of the gas, $v$ the rotational velocity
of the binary-disc system, $v_{\rm bin}$ the keplerian velocity,
$a_{\rm bin}$ the separation of the binary and $h$ the scale height
of the disc. We represent the tidal gap/cavity opening criterion
in the space of parameter $\left(\,(v_{\rm bin}/v)^2\,,\,(c_{\rm s}/v)(h/a_{\rm bin})\right)$
as a solid blue line in Fig.~\ref{GOC}.

The interpretation of Fig.~\ref{GOC} is simple; in the absence of
AGN feedback any combination of parameters above the solid blue
line represents binary-disc systems where no tidal gap/cavity formation
will occur (fast shrinking binaries) and any combination of parameters
below the solid blue line represents binary-disc systems where a tidal gap/cavity
will form (slowly shrinking binaries).
Therefore, if we want to study the effect of AGN feedback
on the migration of a system where a slow/fast migration is expected
we have to select a point below/above the solid blue line and this point
will have the information of the combination of parameters of that
binary-disc system.

In addition to this, we set another constraint to the parameters of
the binary-disc systems. We limit our study to systems that are
stable against self-gravity to avoid the presence of clumps that 
can perturb the orbit of the binary through close encounters,
because their effect on the dynamics of the SMBHB could be
difficult to disentangle from the effect of AGN feedback.

To make this selection clear, we write the Toomre parameter $Q$
\citep{Toomre} as a function of the parameters of Fig.~\ref{GOC}.
For this we take the usual form of the Toomre parameter:

\begin{eqnarray}
  Q&=&\frac{c_{\rm s} k}{\pi G \Sigma}\,,
\end{eqnarray}
where $k$ is the epicylic frequency and $\Sigma$ the
surface density of the disc. We express $k$ as a function
of the orbital frequency $\Omega=v/(\pi\,a_{\rm bin})$ as
$k=4\,\Omega^2+\Omega\,a_{\rm bin}\,(d\Omega/dr)$ which
leads to

\begin{eqnarray}
  k^2&=&\frac{v^2}{a_{\rm bin}^2}\,\frac{1}{\pi^2}\,\left(8\pi+\left[5f-3\right]-6\right)
\end{eqnarray}
with $f=1+(M_{\rm bulge}/M_{\rm bin})$, where $M_{\rm bin}$ is the mass
of the binary and $M_{\rm bulge}$ the enclosed mass of a bulge
of stars, following a Plummer profile, that surrounds
our binary-disc system. The radius at which we compute
this enclosed mass is the orbital radius of the binary,
therefore $f$ depends on the initial separation of the binary.
The values of $f$, for the four binary separations that we
use, range from $f=1.001$ to $f=1.5$. Henceforth
we use its mean value $f\approx1.1$.

We express the surface density of the disc 
as $\Sigma=4\,M_{\rm gas}/(\pi\, a_{\rm bin}^2)$ with $M_{\rm gas}$
the mass of  gas enclosed by the binary.
We express this mass as a function of the velocities of
the system as $M_{\rm gas}=(2\,G/a_{\rm bin})\left(v^2-f\,v_{\rm bin}^2\right)$.

With these expressions for $k$ and $\Sigma$ we rewrite the Toomre stability
parameter as:

\begin{eqnarray}
  Q&=&\frac{(x\,f)^{-1}}{(x\,f)^{-1}-1}\,\sqrt{\frac{A\,y\,\beta}{2\,\pi^2}}\,,
\end{eqnarray}
where $\beta=(c_{\rm s}/v)(a_{\rm bin}/h)$ is a parameter
of order unity for rotational supported systems,
$x=(v_{\rm bin}/v_{\rm gas})^2$, $y=(c_{\rm s}/v_{\rm gas})^2$ and
$A=A_0+x\,A_1$ with $A_0=4\,\pi-3$ and $A_1=(5\,f-3)/2$.

Therefore, the curve that defines $Q=1$ in the parameter space
$\left((v_{\rm bin}/v_{\rm gas})^2\,,\,(c_{\rm s}/v_{\rm gas})(h/a_{\rm bin})\right)$ is

\begin{eqnarray}
 y&=&\frac{2\pi^2}{\beta}\,\frac{(1-x\,f)^2}{A_0+x\,A_1}\,.
\end{eqnarray}

 This curve is shown as a red dashed line in Fig.~\ref{GOC}. The parameter space that we
 use for the initial conditions of the simulations of this paper is at
 the right of the  red dashed line that corresponds to systems with $Q>1$.

\section{SIMULATIONS}
\subsection{Implementation of AGN Feedback}

To study the effect of AGN feedback on the dynamical evolution
of a SMBHB embedded in a circumbinary disc we use a modified
version of the code Gadget-3 \citep{Springel2005} in which we
implement flux accretion on to the SMBHs and the
subsequent AGN radiative/mechanical feedback following \cite{Choi2012}.

 We choose flux accretion instead of Bondi-Hoyle accretion because
  the Bondi radius is typically larger than the size of the binary+disc
  system in our setup.
In our implementation of accretion we define a spherical region
around the SMBH of radius $R_{\rm acc}$ in which all the gas that enters
is eligible to be accreted by the SMBH. In addition the gas has to fulfill an angular momentum condition;
the gas will be accreted by the SMBH only if its angular momentum
is smaller than the angular momentum of a circular orbit of radius
$R_{\rm ss, disc}$. We define $R_{\rm ss, disc}$ as the maximum radius
of a Shakura and Sunyaev disc that is stable against self gravity
\citep{KolSyun1980}. The effect of this condition on the accretion
rate is presented in appendix \ref{AccretionApp}.

 To model radiative feedback we compute the total radiative
  luminosity emitted by the SMBH accreting at a rate $\dot{M}_{\rm BH}$
  as $L_{\rm r}=\epsilon_{\rm r}\dot{M}_{\rm BH} c^2$ where $c$ is the speed of
light and $\epsilon_{\rm r}$ is the radiative efficiency that, following 
\citet{Choi2012}, we set as $\epsilon_{\rm r}=0.1$.
This total luminosity is used to compute an X-ray feedback
(coming from Compton cooling/heating, photo-ionization and electron scattering) and an Eddington force (coming from
electron scattering). For the X-ray feedback we convert the
total radiated luminosity to a luminosity flux $F_{\rm r}=L_{\rm r}/ 4 \pi r^2$
, where $r$ denotes the distance of the gas particles
to the SMBH. Then we convert this flux in a net volume
heating rate term  ${\dot E}$ using the description of \cite{sazonov} of the net heating rate
per unit volume of a cosmic plasma in photoionization equilibrium embedded
in a radiation field that corresponds to the average spectral
energy distribution of quasars \citep[e.g.][]{novak}.
We use $\dot E$ to heat every gas particle and we include the effect
of X-ray pressure by adding a total momentum per unit time
${\dot p}={\dot E}/c$ \citep{debuhr2011,debuhr2012} which does not
take into account the effect of dust.
For the Eddington force we use the luminosity flux $F_{\rm r}$ to compute the total momentum
per unit time on to each gas particle due to electron scattering as
${\dot p}=F_{\rm r}\,N_{e}\,\sigma_{\rm T}/c$ with $N_{e}$
the number of free electrons and $\sigma_{\rm T}$ the Thomson cross-section
for electrons.

The mechanical part of the AGN feedback models line-driven winds launched in accretion discs \citep{2001AdSpR..28..459P}.
To model this wind, as in \citet{Choi2012}, we assume 
a single-velocity wind with  $v_{\rm w}=10^4$ km s$^{-1}$ and we define
the wind mass rate as $\dot{M}_{\rm w}$. Defining $\dot{M}_{\rm inf}$
as the rate of gas mass that fulfills the conditions of flux
accretion then, by simple mass conservation, we have that
$\dot{M}_{\rm BH}=\dot{M}_{\rm inf}-\dot{M}_{\rm w}$. Therefore, defining
  $\phi=\dot{M}_{\rm w}/\dot{M}_{\rm BH}$ we can write the SMBH
accretion rate, the wind mass rate and the momentum and kinetic
energy of the wind as
\begin{eqnarray}
  \dot{M}_{\rm BH}&=&\dot{M}_{\rm inf}\,\frac{1}{1+\phi} \\
  \dot{M}_{\rm w}&=&\dot{M}_{\rm inf}\,\frac{\phi}{1+\phi} \\
  \dot{E}_{w}&=&\epsilon_{\rm w}\,c^2\,\dot{M}_{\rm inf}\,\frac{1}{1+\phi} \\ 
  \dot{p}_{w}&=&\dot{M}_{\rm inf}\,v_{\rm w}\,\frac{\phi}{1+\phi}\,,
\end{eqnarray} 
where $\epsilon_{\rm w}=5\times10^{-4}$ denotes the wind efficiency
in the momentum feedback model. Using these equations we can write $\phi=2\epsilon_{\rm w}c^2/v_{\rm w}^2\approx 0.9$ which give us
$\dot{M}_{\rm w}\approx 0.5\,\dot{M}_{\rm inf}.$
Therefore, to model this wind we select half of the particles that are eligible to
be accreted by the SMBH and we eject them with a velocity $v_{\rm w}$
radially from the SMBH in the direction of the original angular
momentum of the particles. In this way we mimic a wind that is
launched perpendicular to the accretion disc.

Finally, we include cooling to model the radiative losses of the gas.
The cooling function that we use is computed for an
optically thin gas with solar metallicity. For temperatures between
$10^{4}$ and $10^{8}$  we compute the cooling function essentially
as described by \citet{Katz1996} (bremsstrahlung and lines cooling)
and for temperatures below $10^{4}$ we use the parametrization of \citet{GI1997}.
For this cooling function, and the mean density of the gas on the initial
conditions, the cooling time is of the order of $t_{\rm cool}\,\approx$ 1 yr, 
which is much shorter than the orbital time of the binary ($\sim$ 1-100 kyr).
For this efficient cooling the energy injected thermally to the gas is radiated
  away in a time $\sim t_{\rm cool}$. Therefore, thermal AGN feedback is
completely ineffective for the scales and densities that we explore in this
study. For this reason we implement a radiative and momentum based AGN
feedback instead of a purely thermal AGN feedback.

\subsection{Initial Conditions and Simulations}

We select the parameters of the binary-disc systems using the
constraints presented in section \ref{ICS_CONTRAINTS}.
We select four binary-disc systems as initial conditions
and they are represented as colored squares in Fig.~\ref{GOC}.
Of these four initial conditions, two correspond
to systems with an initial tidal gap/cavity (label
with ``GAP'' in table \ref{ics_table}).

In the four binary-disc systems that we generate the gas disc
follows a Mestel surface density profile with a maximum
radius of $R_{\rm disc}=45$ pc, an initial thickness of $H_{\rm disc}=5$ pc
{\bf and a uniform gas temperature of $2\times10^4$ K.}
In three of these initial conditions the total mass of the disc is $10^6\,M_{\odot}$
and in one of them is $10^{4}\,M_{\odot}$.
We model these discs using $2\times10^6$ SPH particles and for one case we
use $2\times 10^5$ SPH particles (labeled as ``lowr'' in table \ref{ics_table}),
each with a gravitational softening of 0.005 pc.

In addition to the disc we include an external
potential to mimic the presence of a spherical bulge of stars,
formed at the core of a galaxy merger remnant, around the
binary-disc system.
We include this external potential because such bulge can have
an impact on the dynamics of the gas launched as a wind from the SMBHs.
We model this bulge as a Plummer potential of core radius
65 pc and total mass $2.5\times 10^{8}\,M_{\odot}$,
consistent with the black hole mass - bulge mass relation
\citep{MarconiHunt2003,HaringRix2004} for a black hole
of $10^6\,M_{\odot}$.

Before adding the binary to the initial conditions
we relax the gaseous disc, for 30 orbits, under the
presence of the Plummer potential and a spherical
external potential that mimics the presence of the binary.
This external potential is a homogeneous sphere of mass
$M_{\rm bin}$ and radius equal to the binary separation
$a_{\rm bin}$.

After the disc is relaxed we select all the gas particles
that are in a ring of radius $a_{\rm bin}/2.0$, and width $0.02\,a_{\rm bin}$,
and we compute their mean rotational velocity.
Then we add an equal mass SMBHB in a circular orbit
with this velocity, mass $M_{\rm bin}$ and separation $a_{\rm bin}$.
The total mass of the binary and its initial separation in
the four initial conditions are
$(a_{\rm bin},M_{\rm bin})\,=\,(1\,{\rm pc},10^6\,M_{\odot}),\,(2\,{\rm pc},10^6\,M_{\odot}),\,(2\,{\rm pc},10^5\,M_{\odot}),\,(3\,{\rm pc},10^4\,M_{\odot})$
For all these cases we model the binary with two collision-less
particles of gravitational softening 0.005 pc.

In table \ref{ics_table} we list the simulations.
For each of the four initial conditions we run two simulations: one without
AGN feedback and one with AGN feedback (label  ``AGN'' in table \ref{ics_table})

\begin{table}
	\centering
	\label{ics_table}
        \caption{Table of simulation parameters.
          $M_{\rm bin}$ is in units of $10^6\,M_{\odot}$ and
          $a_{\rm bin}$ in units of parsec.
          The ``lowr'' label refers to a simulation with
          $2\times 10^5$ gas particles instead of
          $2\times 10^6$.
        }
	\begin{tabular}{lcccccc} 
		\hline
		Label &$M_{\rm bin}$ &$M_{\rm disc}/M_{\rm bin}$ & $a_{\rm bin}$ & Tidal Gap & AGN\\
		\hline
		{\color{darkblue} GAP-a3}        & 1    & 0.01 & 3  & yes &  no \\
                {\color{darkblue} GAP-a3-AGN}    & 1    & 0.01 & 3  & yes &  yes \\
		{\color{Dgreen} GAP-a2}          & 1    & 1    & 2  & yes &  no   \\
                {\color{Dgreen} GAP-a2-AGN}      & 1    & 1    & 2  & yes &  yes  \\
                {\color{brownRed} a2}            & 0.1    & 10   & 2  & no  &  no   \\
                {\color{brownRed} a2-AGN}        & 0.1    & 10   & 2  & no  &  yes  \\
                {\color{orange} a1-lowr}         & 0.01 & 100  & 1  & no  &  no   \\
                {\color{orange} a1-lowr-AGN}     & 0.01 & 100  & 1  & no  &  yes  \\
                {\color{orange} a1 }             & 0.01 & 100  & 1  & no  &  no   \\
                {\color{orange} a1-AGN}          & 0.01 & 100  & 1  & no  &  yes  \\

		\hline
        \end{tabular}
\end{table}

\section{RESULTS}
\subsection{Evolution of SMBH binaries without AGN feedback}

\begin{figure}
	\includegraphics[width=\columnwidth]{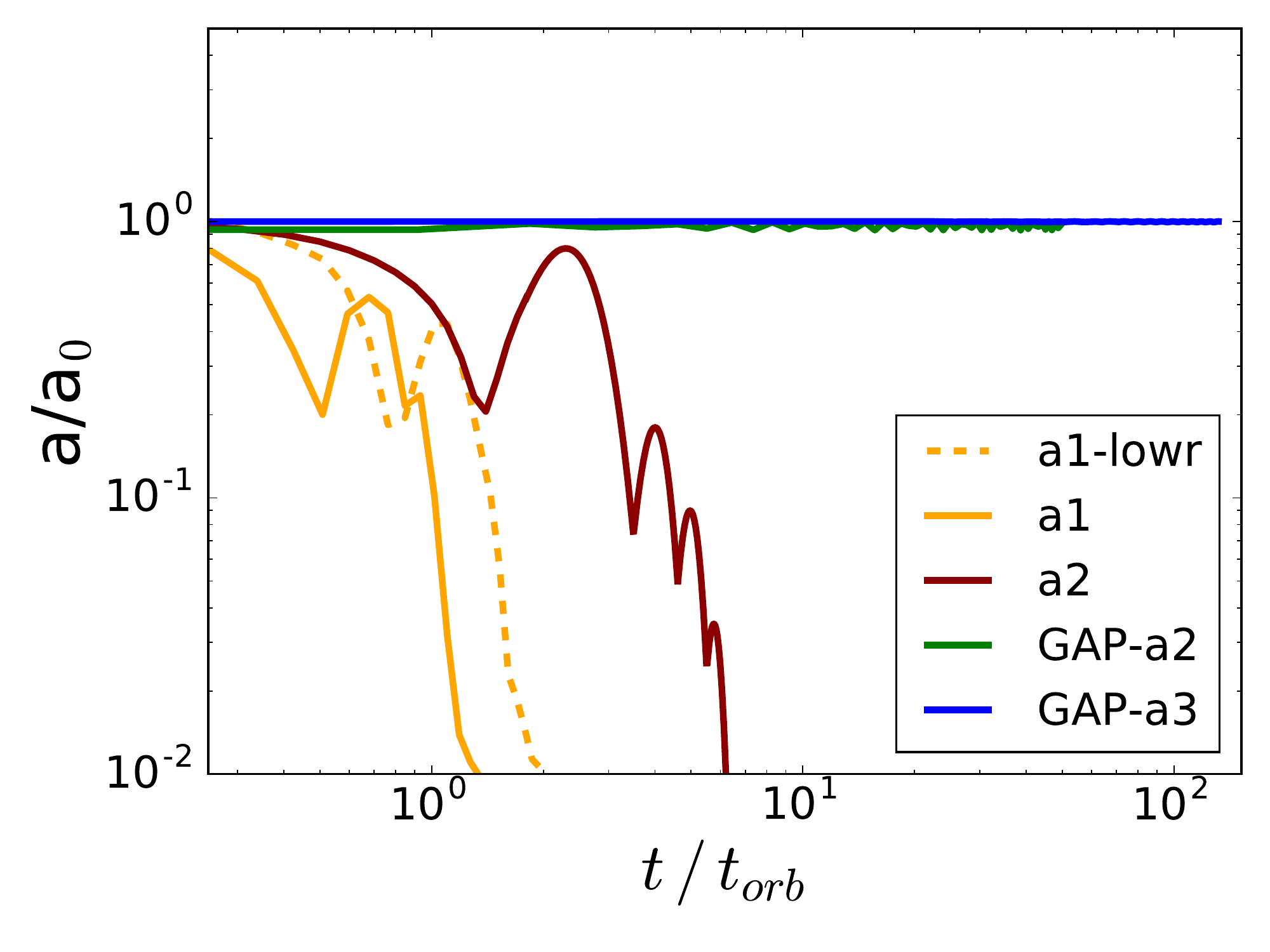}
        \caption{Separation of the binary as a function of time for simulations
          without AGN feedback. The binary separation
          $a_{\rm bin}$ is normalized by the initial binary separation $a_{0}$ and the time
          is in units of the initial orbital time $t_{\rm orb}$.
          In simulations where a tidal cavity forms
          ({\color{Dgreen} GAP-a2} and {\color{darkblue}  GAP-a3})
          the binary separation is almost constant with time (slow shrinking)
          while in simulations where no tidal cavity is formed
          ({\color{orange} a1}, {\color{orange} a1-lowr} and {\color{brownRed} a2})
          the binary separation evolves rapidly.
          }
    \label{DIST_NOAGN}
\end{figure}

In Fig.~\ref{DIST_NOAGN} we show the temporal evolution of 
the binary separation $a_{\rm bin}$ normalized to the initial binary separation
$a_0$ for simulations without AGN feedback. 

In the simulations {\color{darkblue}  GAP-a3}
and {\color{Dgreen} GAP-a2} the binary separation is constant over a
timescale of roughly $\sim\,10^2\,t_{\rm orb}$, which is consistent with
the expectations of the migration timescale of a system with tidal cavity (decrease on a timescale of
$(da_{\rm bin}/dt)\,(1/a_{\rm bin})\sim\,10^{3}-10^{4}\,t_{\rm orb}$).

Also consistent with the analytic expectations, in the simulations {\color{orange} a1} 
and {\color{orange} a1-lowr} the binary shrinks in $\sim\,1-2\,t_{\rm orb}$ which
is consistent with a system in which the redistribution of angular momentum on the disc
is rapid enough to maintain gas close to the binary and therefore where no tidal cavity
is formed. 


\subsection{Evolution of SMBH binaries with AGN feedback}

\begin{figure}
	\includegraphics[width=\columnwidth]{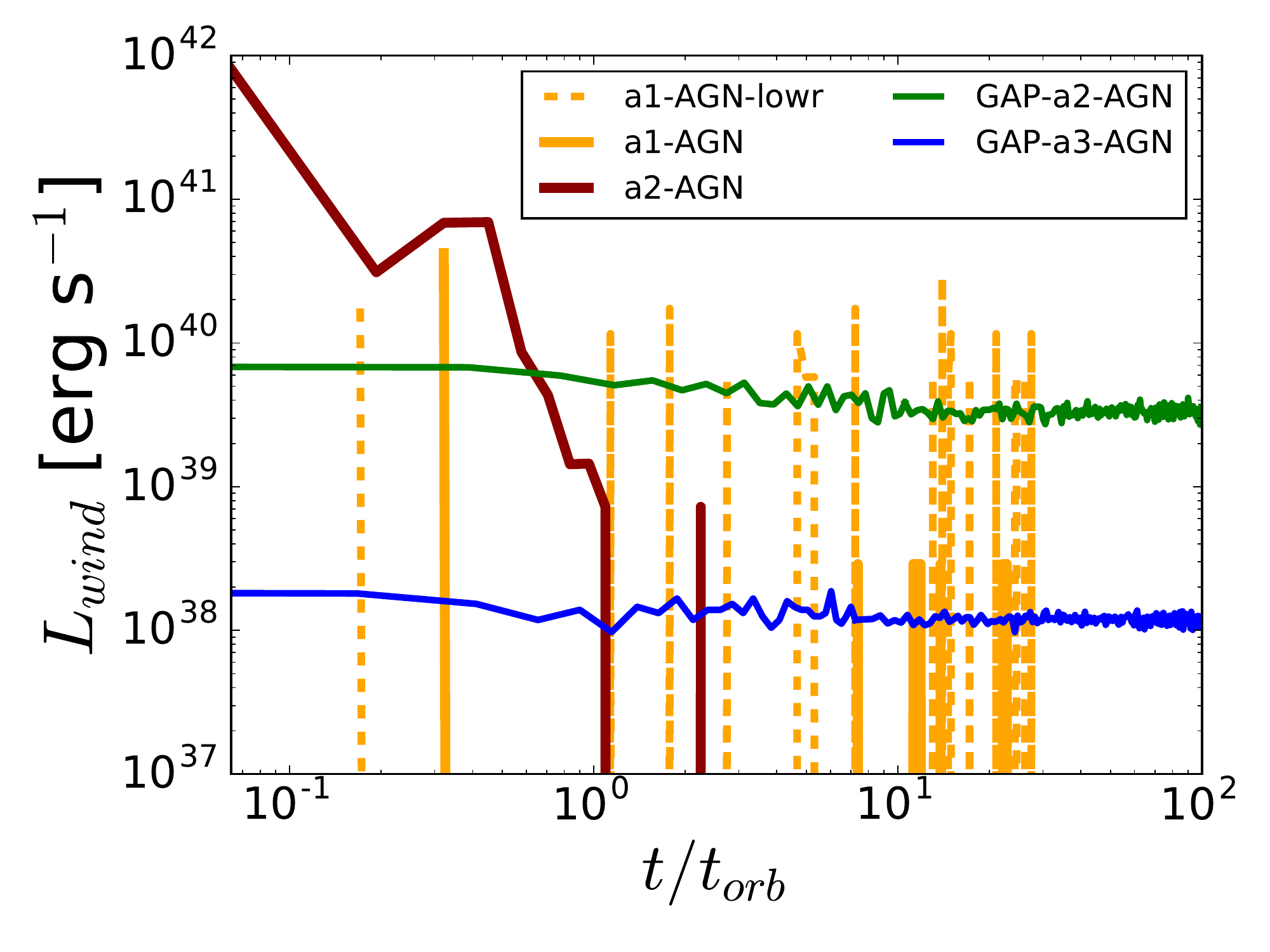}
        \caption{Luminosity of the wind launched from the SMBHB
          as a function of time.
          The time is in units of the initial orbital time $t_{\rm orb}$.
          Even though in simulations without
          tidal cavity ({\color{orange} a1-lowr-AGN},
          {\color{orange} a1-AGN} and {\color{brownRed} a2-AGN})         
          the mass of the black holes are smaller, the luminosity of the wind reaches higher
          values than in simulations with tidal cavity
          ({\color{darkblue} GAP-a3-AGN} and {\color{Dgreen} GAP-a2-AGN})         
          where the mass of the black holes is $\sim 10-100$ times larger.
        }
        \label{WIND_LUM}
\end{figure}
\begin{figure}
  \includegraphics[width=\columnwidth]{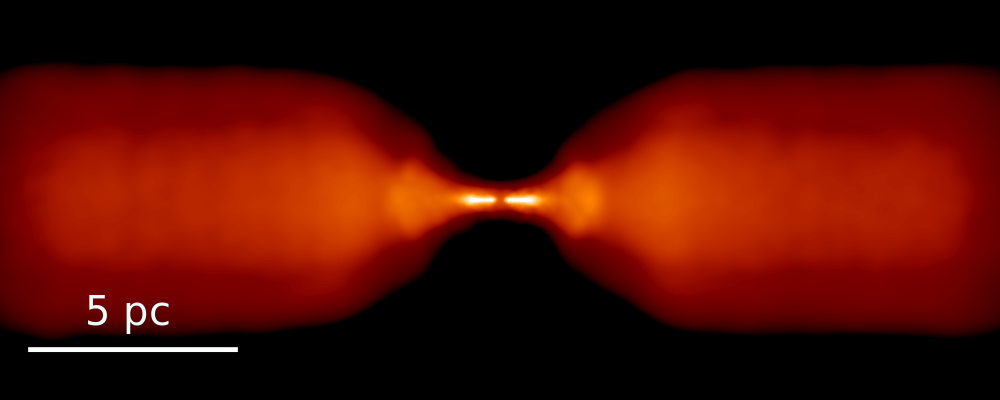}\\
  \includegraphics[width=\columnwidth]{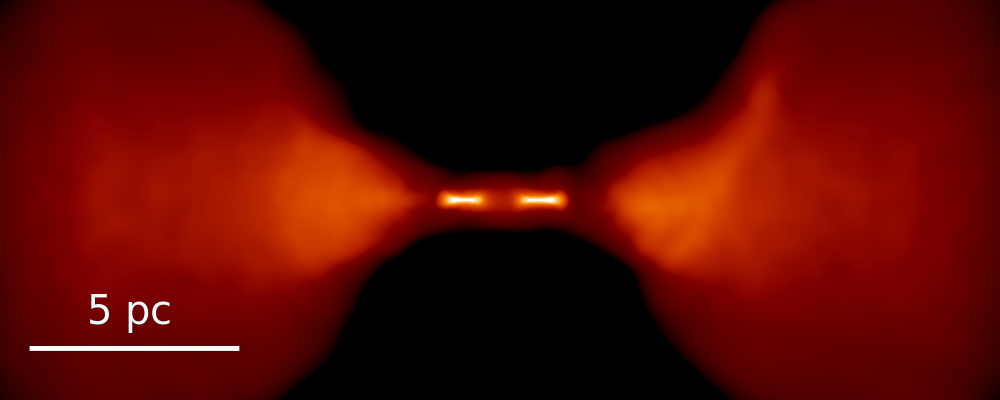}\\
  \includegraphics[width=\columnwidth]{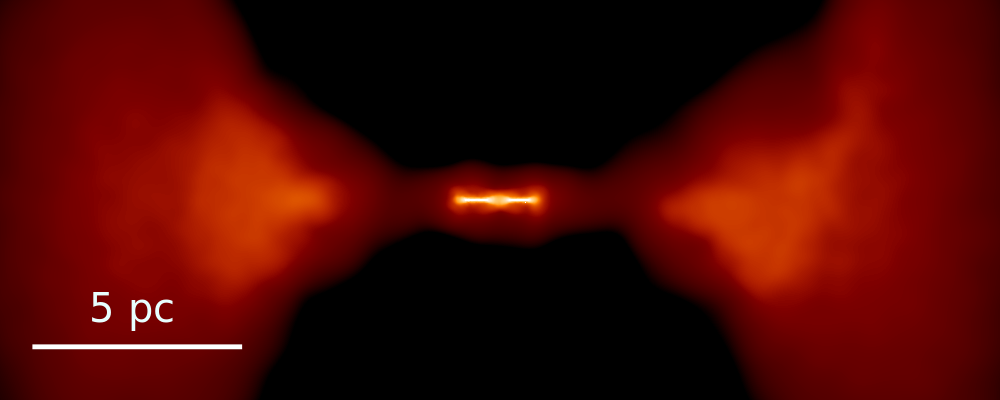}\\
  \includegraphics[width=\columnwidth]{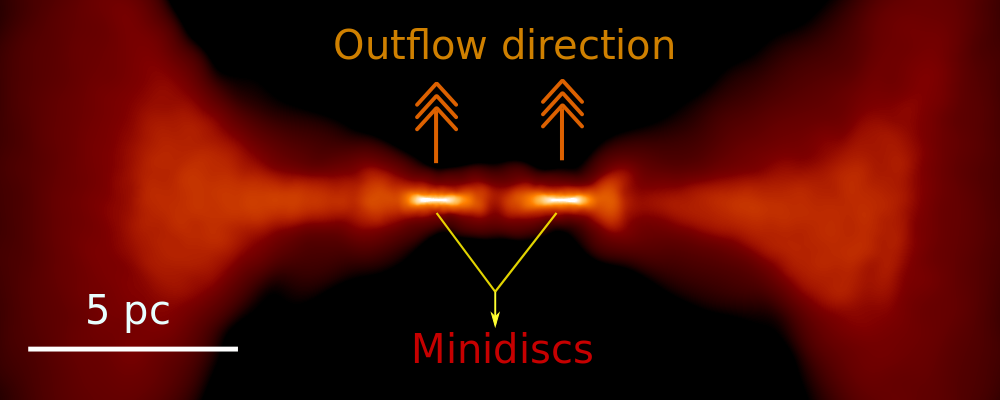}
  \caption{Edge-on density slices of the disc in simulation {\color{darkblue}GAP-a3-AGN}.
    From top to bottom each panel corresponds to a different time.
    The presence of a cavity and the alignment of the
    mini-discs, from where the SMBHs accrete, with the circumbinary disc
    leaves a free channel in the direction of the wind. Therefore,
    the winds launched from the SMBHBs, perpendicular to the mini-discs,
    can escape without perturbing the circumbinary disc. 
    }
    \label{AGN_GAP_A3_EDGEON}
\end{figure}

In the presence of a tidal cavity the 
inflow of gas towards the SMBHs is smaller than
in simulations without tidal cavity and therefore
the maximum wind luminosity is lower (see Fig.~\ref{WIND_LUM}).
Furthermore, the outflows do not interact with the disc. They escape
freely through the cavity and their velocity does not change, leaving the disc almost
unperturbed (see Fig.~\ref{AGN_GAP_A3_EDGEON}).

\begin{figure}
	\includegraphics[width=\columnwidth]{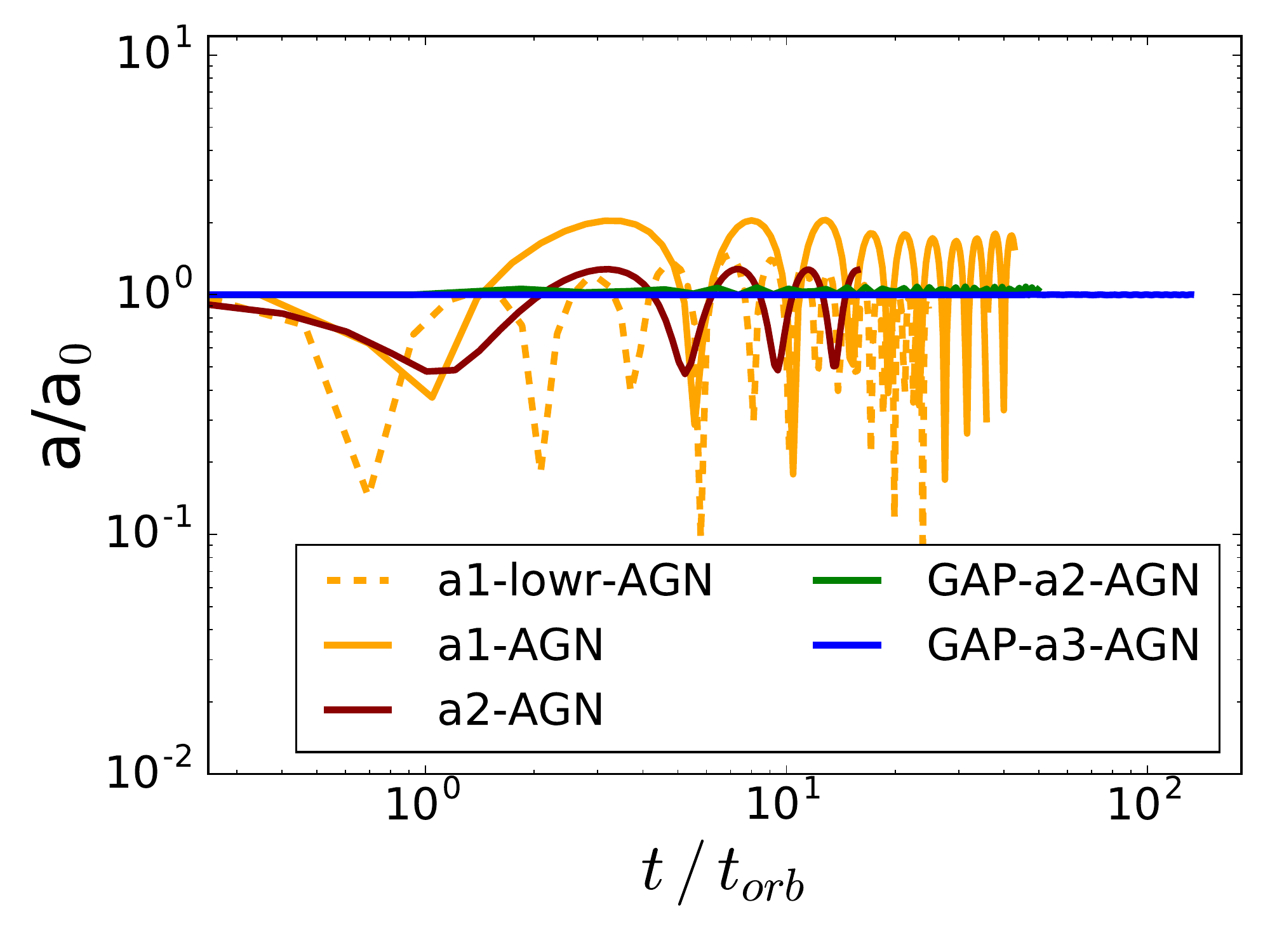}
        \caption{Separation of the binary as a function of time for simulations
          where accretion and AGN feedback of the SMBHs is active.
          The binary separation $a_{\rm bin}$ is normalised by the initial binary
          separation $a_{0}$ and the time is in units of the initial orbital time $t_{\rm orb}$.
          The evolution of systems with tidal cavity
          is not affected by  AGN feedback ({\color{darkblue} GAP-a3-AGN} and
          {\color{Dgreen} GAP-a2-AGN}). However, in simulations without tidal cavity
          ({\color{orange} a1-lowr-AGN}, {\color{orange} a1-AGN} and {\color{brownRed} a2-AGN})
           AGN feedback does not allow the binary to shrink further.
          }
    \label{DIST_AGN}
\end{figure}

This means that when a tidal gap opens, the  binary separation for simulations with
AGN feedback ({\color{darkblue} GAP-a3-AGN} and
{\color{Dgreen} GAP-a2-AGN} in Fig.~\ref{DIST_AGN}) has the
same behaviour that their counterpart in simulations without
AGN feedback ({\color{darkblue} GAP-a3} and {\color{Dgreen} GAP-a2},
in Fig.~\ref{DIST_NOAGN}). 

\begin{figure}
  \includegraphics[width=\columnwidth]{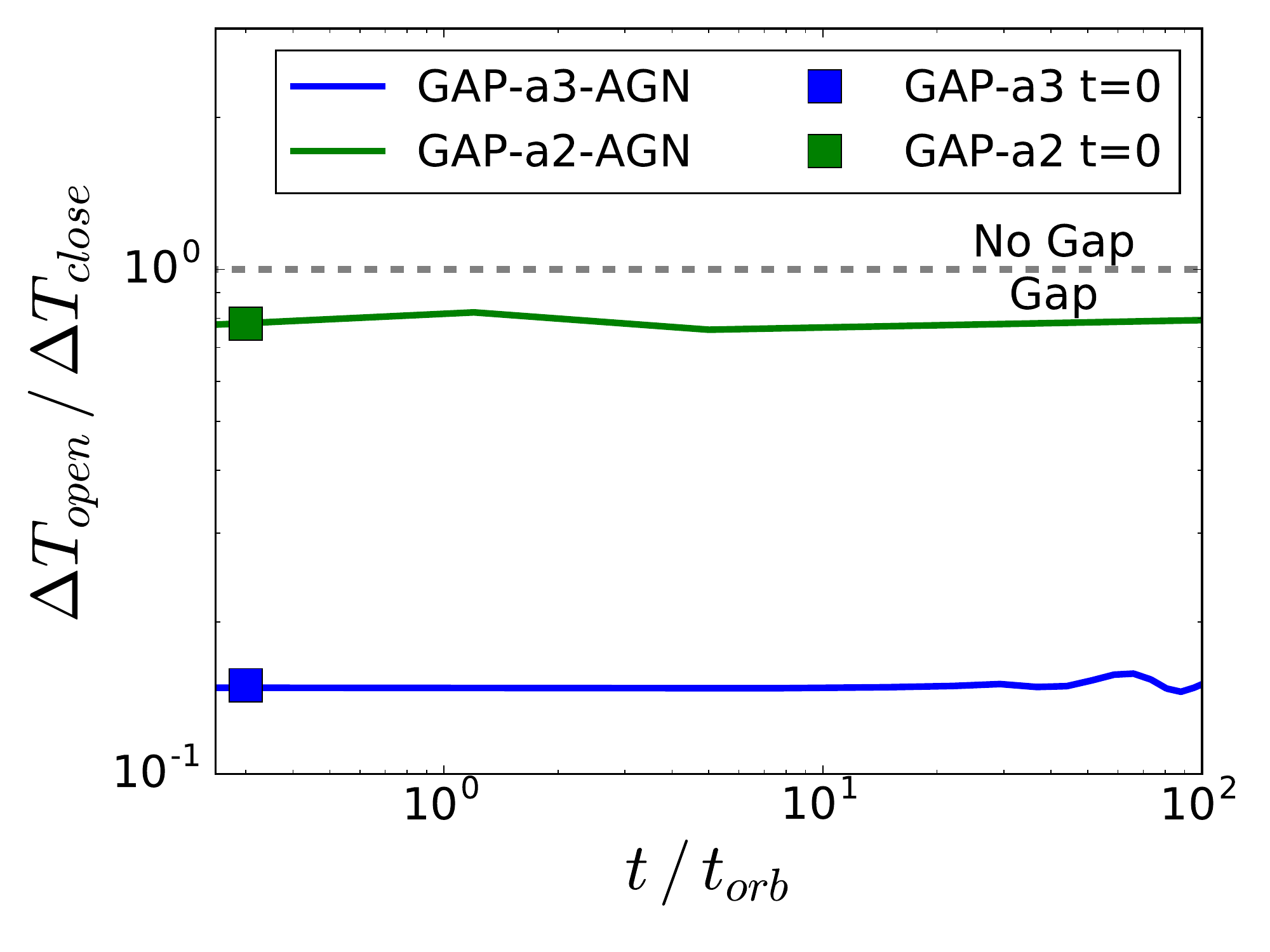}
  \caption{Time evolution of $\Delta t_{\rm open}/\Delta t_{\rm close}$
    after the onset of AGN feedback for simulations
    with a tidal cavity. The dashed grey line
    corresponds to $\Delta t_{\rm open}/\Delta t_{\rm close}=1$
    and represents the threshold between fast migration
    and slow migration regime.
    The blue line corresponds to simulation {\color{darkblue} GAP-a3-AGN}
    and the green line to simulation {\color{Dgreen} GAP-a2-AGN}.
    The squares represent the initial value of
    $\Delta t_{\rm open}/\Delta t_{\rm close}$ for both simulations.
    The value of $\Delta t_{\rm open}/\Delta t_{\rm close}$ for both simulations
    does not change in time because the
    BHs accretion is low and the winds generated by the AGN
    feedback escape from the circumbinary disc without altering it.
    }
    \label{GOC_time}
\end{figure}

\begin{figure}
  \includegraphics[width=\columnwidth]{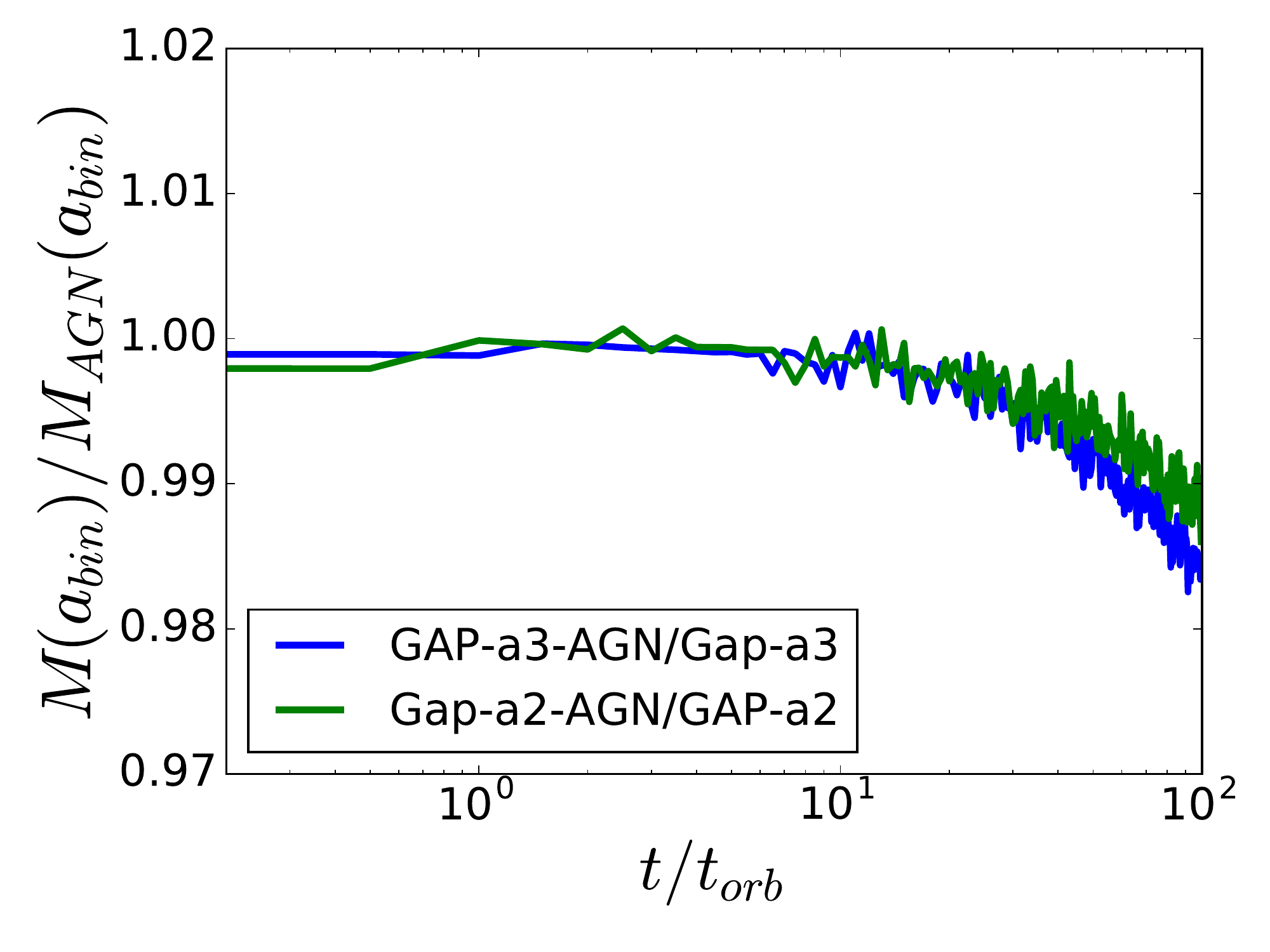}
  \caption{Ratio between the mass of gas inside $r=a_{\rm bin}$
    for simulations with and without AGN feedback as a function of time.
    The blue line corresponds to the ratio of the enclosed mass of simulations
    {\color{darkblue} GAP-a3-AGN} and {\color{darkblue} GAP-a3}.
    The green line corresponds to the ratio of the enclosed mass of simulations
    {\color{Dgreen} GAP-a2-AGN} and {\color{Dgreen} GAP-a2}.
    Time is in units of the initial orbital time $t_{\rm orb}$.
    The enclosed mass in both binary-disc
    systems with cavity changes by 1-2 \% by the effect of AGN feedback.
    This results in a negligible effect of AGN feedback on the
    torques produced by the disc on to the SMBHBs.}
    \label{MABIN_A3}
\end{figure}

In these two simulations with AGN feedback the amount of matter in the outflows is very
small compared with the mass of the disc enclosed by the orbit
of the binary (Fig.~\ref{MABIN_A3}).
After $\sim 10^2\,t_{\rm orb}$ the enclosed mass is about
the same for simulations {\color{darkblue} GAP-a3-AGN} and
{\color{darkblue} GAP-a3} and for simulations
{\color{Dgreen} GAP-a2} and {\color{Dgreen} GAP-a2-AGN}.
This implies that the effect of AGN feedback on the orbit of a binary
in a disc with tidal cavity is negligible. Moreover,
the discs in both pairs of simulations are basically indistinguishable.
 Because the discs are basically indistinguishable, simulations
  {\color{darkblue} GAP-a3-AGN} and {\color{Dgreen} GAP-a2-AGN} 
 remain in the slow migration regime (see figure \ref{GOC_time}).

In simulations that are not expected to form a tidal cavity
without AGN feedback, the presence of this feedback has a strong effect on
the dynamical evolution of the binary and the structure of the disc. 
When a cavity does not open, SMBHs are surrounded by dense gas and large  
inflows can reach the BHs triggering strong winds, as shown in Fig.~\ref{WIND_LUM}.

\begin{figure}
  \includegraphics[width=\columnwidth]{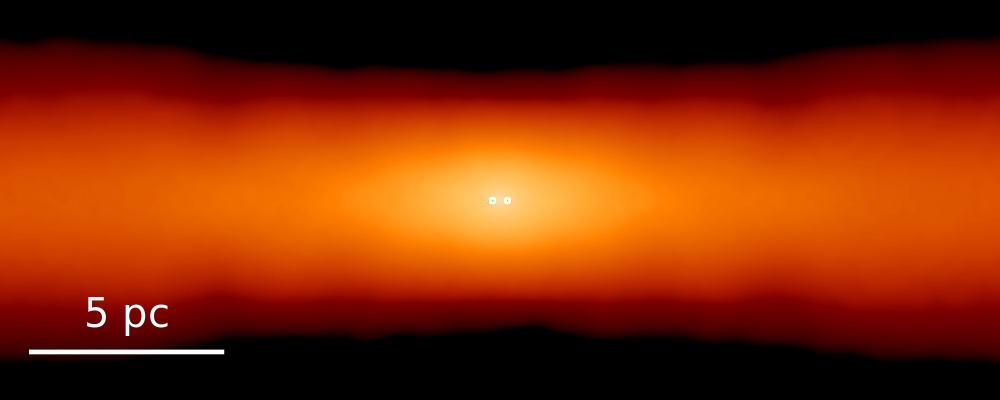}\\
  \includegraphics[width=\columnwidth]{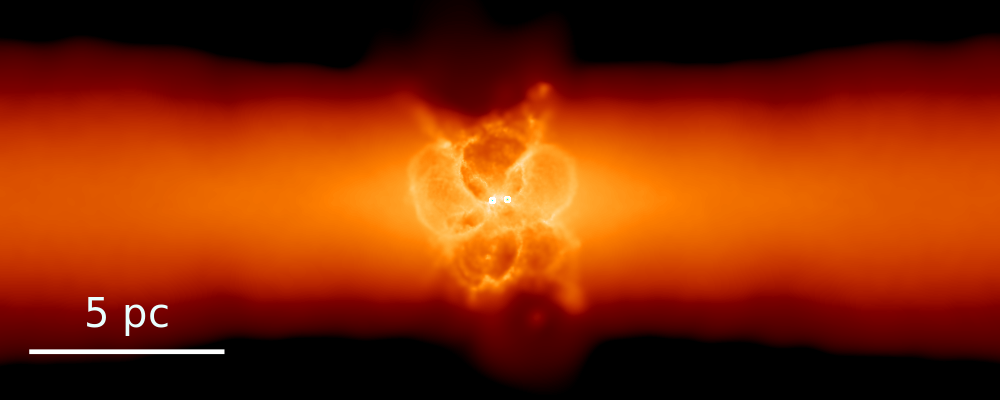}\\ 
  \includegraphics[width=\columnwidth]{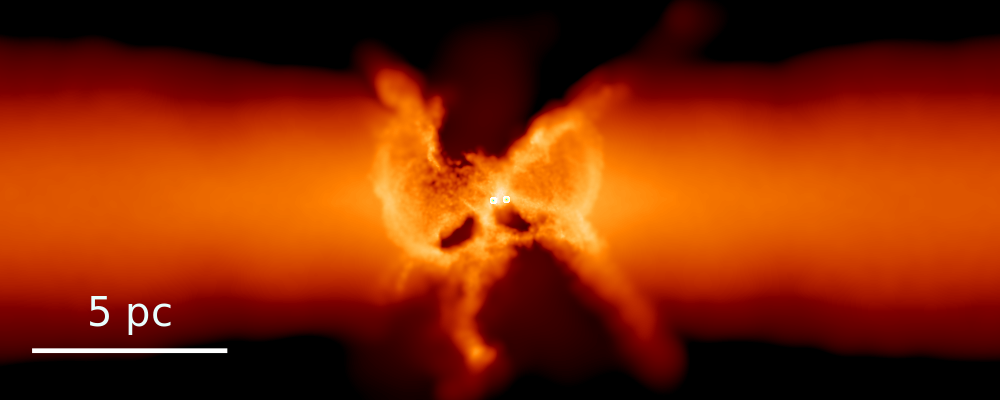}\\
  \includegraphics[width=\columnwidth]{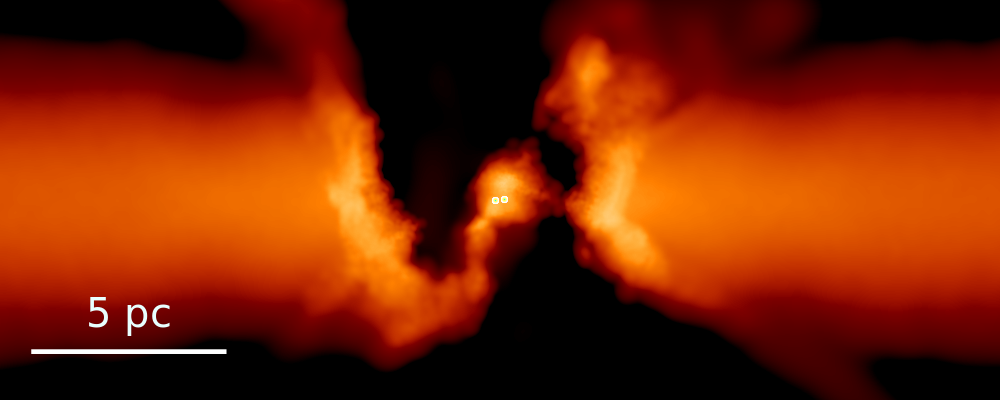}
  \caption{Edge-on density slices of the disc in simulation {\color{orange}a1-AGN}.
    From top to bottom each panel corresponds to a different time.
    AGN winds collide against the disc, opening
    a feedback cavity which leaves both SMBHs almost naked. This results in
    the subsequent stalling of the SMBHB migration.
  }
    \label{AGN_A1_EDGEON}
\end{figure}

\begin{figure}
  \includegraphics[width=\columnwidth]{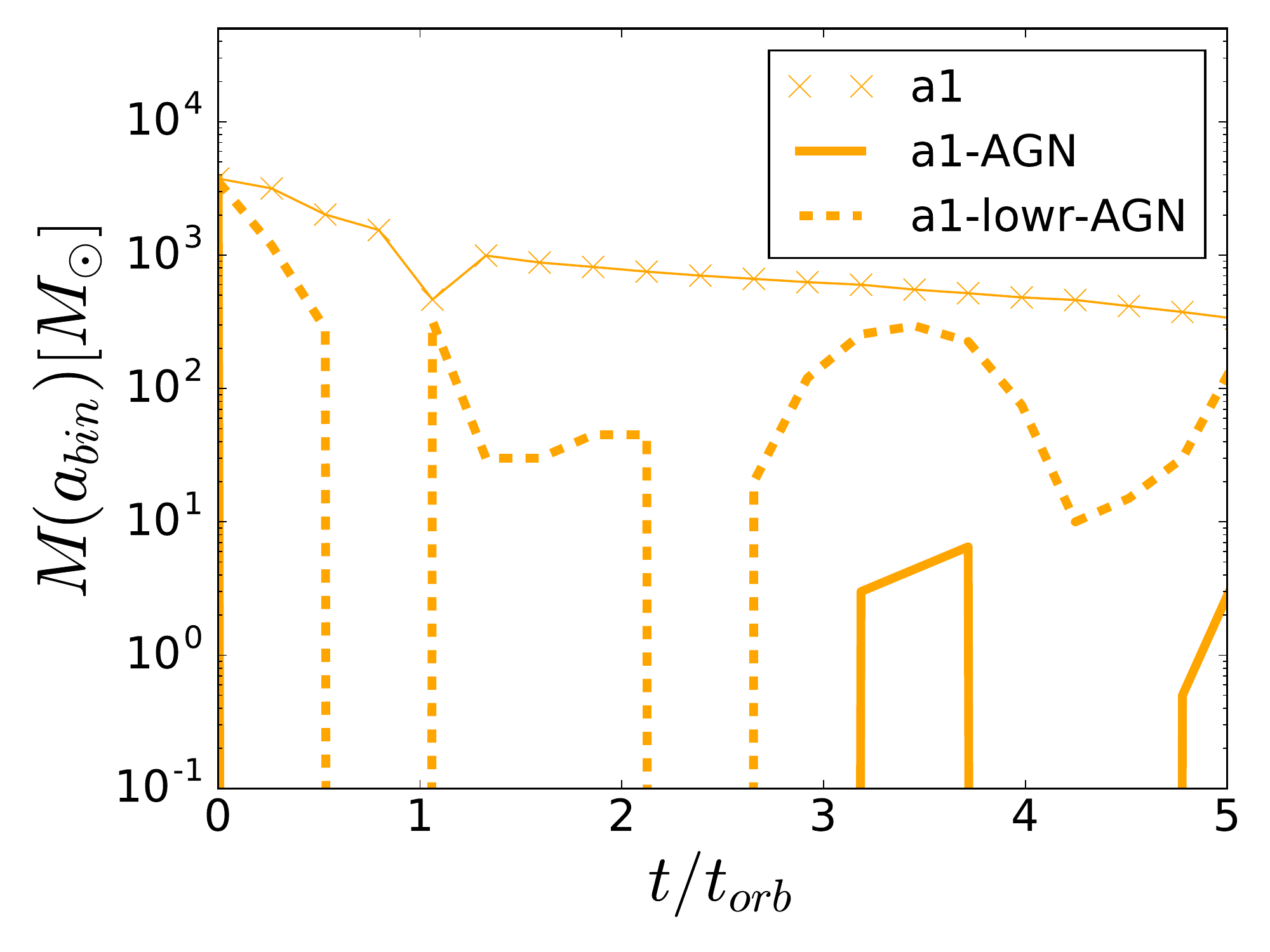}
  \caption{Mass of gas inside $r=a_{\rm bin}$ as a function of time
    for three simulations: {\color{orange}a1} (thin crossed line), {\color{orange}a1-AGN} (dashed thick line)
    and {\color{orange}a1-lowr-AGN} (solid thick line).
    The enclosed mass is in units of $M_{\odot}$ and time is in units of the
    initial orbital time $t_{\rm orb}$.
    AGN feedback excavates a ``feedback cavity'',
    decreasing the enclosed mass, on different timescales
    for simulation {\color{orange}a1-AGN} and {\color{orange}a1-lowr-AGN}.
    The binary therefore stalls at different separations in both simulations.
  }
  
    \label{MABIN_A1}
\end{figure}

In the simulations {\color{orange} a1-lowr-AGN}, {\color{orange} a1-AGN}
and {\color{brownRed} a2-AGN} the SMBHBs do not evolve as rapidly as
they do in the absence of AGN feedback (Fig.~\ref{DIST_NOAGN}),
instead they stall at parsec scale separations (Fig.~\ref{DIST_AGN}).
This stalling is caused by the decrease of mass around the
binary due to the interaction of strong AGN winds with the disc.
This is evident in the time evolution of an edge-on view of the disc in
the simulation {\color{orange} a1-AGN} shown in Fig.~\ref{AGN_A1_EDGEON}.
The AGN wind, without a restricted direction as in simulation
{\color{darkblue} GAP-a3-AGN} and without a tidal cavity to escape,
crashes on to the disc pushing matter away from the binary, thus creating
a ``feedback cavity''.
The same can be seen in Fig.~\ref{MABIN_A1} where we show
how the mass around the binary drastically decreases in less than
one orbit for simulations with AGN feedback ({\color{orange} a1-AGN} and
{\color{orange} a1-lowr-AGN}) and slowly decreases in the simulation
without BH accretion and AGN feedback {\color{orange} a1}.

 The formation of this ``feedback cavity'' cannot be predicted
  by the gap opening criterion, because this criterion is derived
  from the competition between gravitational torques and viscous diffusion. Instead, 
  the opening mechanism for this ``feedback cavity'' is the momentum injected
  by the AGN, not the gravitational torques produced by the binary.

Although in simulations {\color{orange} a1-AGN} and
{\color{orange} a1-lowr-AGN} (that only differ in
the mass resolution of the disc) the presence accretion/feedback
results in the stalling of the binary, the evolution of the binary separation is not
exactly the same (see Fig.~ \ref{DIST_AGN}).
This difference is produced by the stochastic implementation
of the AGN wind that results in a different time evolution of
the mass around the binary. In simulation {\color{orange} a1-lowr-AGN}
the formation of a ``feedback cavity'' takes about half an
orbit more than in simulation {\color{orange} a1-AGN} (Fig.~\ref{MABIN_A1}).
The binary in simulation {\color{orange} a1-lowr-AGN} has more time
to shrink through tidal torques before the AGN winds
push away all the gas around the binary, the binary the binary
in simulation {\color{orange} a1-lowr-AGN} stalls at a smaller
separation than the binary in simulation {\color{orange} a1-AGN},
as shown in Fig.~\ref{DIST_AGN}.

Regardless of this difference we conclude that in systems for which
in the absence of AGN feedback we expect a fast shrinking,
the effect of the AGN feedback is to delay the shrinking of the SMBHB
by pushing away the gas in which the binary is initially embedded.

\subsection{Effect of AGN feedback on the circumbinary disc}

AGN feedback also has an impact on the circumbinary disc.
  As it was the case for the binary separation, the effect is more
  noticeable for simulations without an initial tidal cavity.

\begin{figure*}
  \includegraphics[width=350pt]{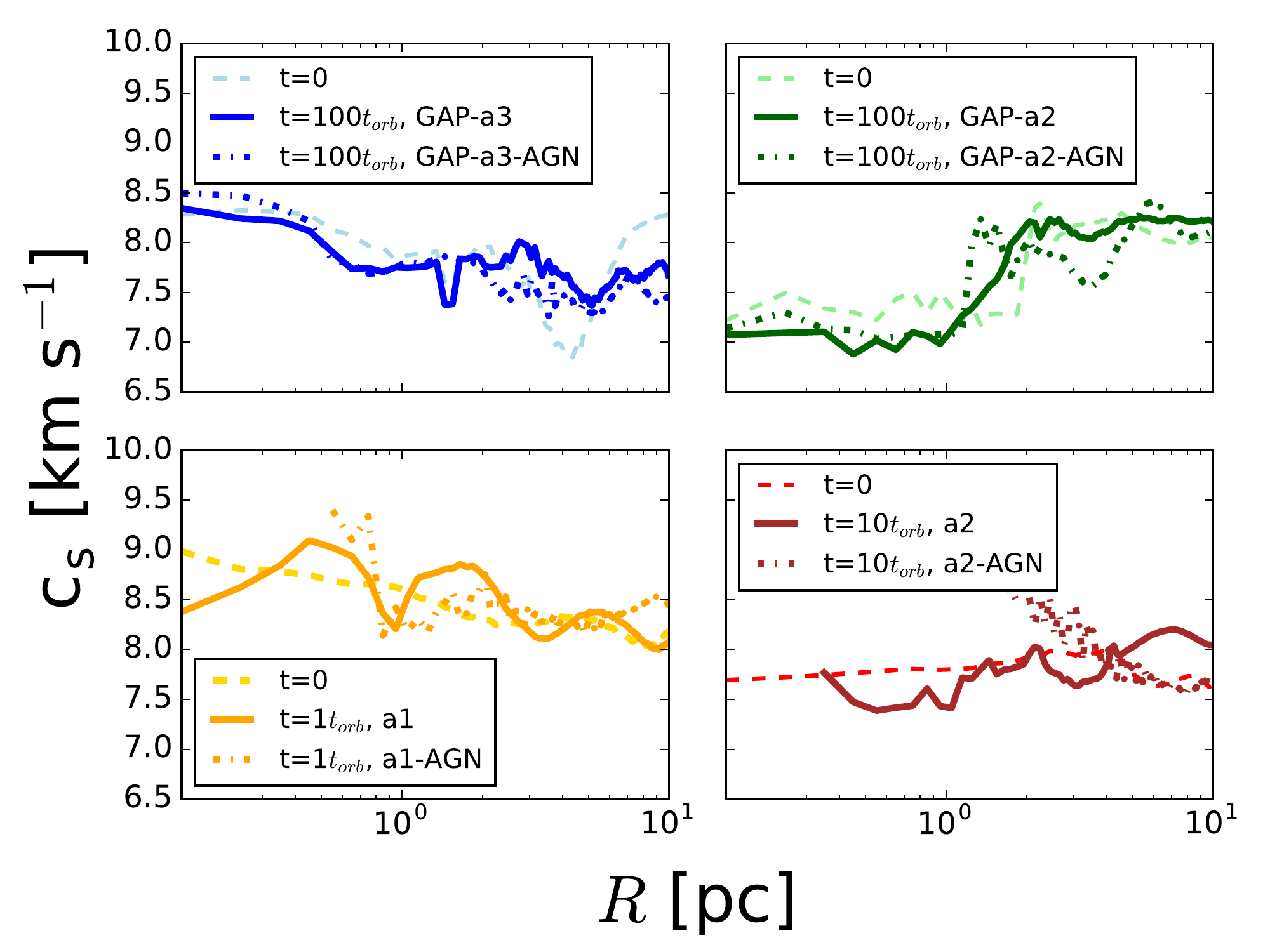}
  \caption{Sound speed of the gas $c_s$ as a function of radius.
    Each panel corresponds to simulations with the same initial conditions.
    The top panels correspond to simulations with tidal cavity
    and the bottom panels to simulations without tidal cavity.
    In each panel $c_s$ is shown at the beginning of the simulation (dashed line), at the end
    of the simulation without AGN feedback (solid line)
    and at the end of the simulation with AGN feedback (dotted line).
    In simulations with an initial tidal cavity/gap (top panels)
    the sound speed is not strongly affected by AGN feedback, however,
    in simulations without an initial tidal cavity/gap (bottom panels)
    AGN feedback opens a ``feedback cavity'' and increases the sound speed of the gas
    in this inner region.
        }
  \label{CsR}
\end{figure*}

For systems with tidal cavity, AGN feedback has almost no
effect on the sound speed of the disc (top panels Fig.~\ref{CsR}).
However, in simulations without a tidal cavity there is an increase
in $c_s$ from $\sim$8.5 km s$^{-1}$ to $\sim$9 km s$^{-1}$ (which corresponds
to a variation on temperature of $\sim$10$^3$ K) at radii close to the edge
of the ``feedback cavity'' (bottom panels Fig.~\ref{CsR}).

\begin{figure*}
  \includegraphics[width=340pt]{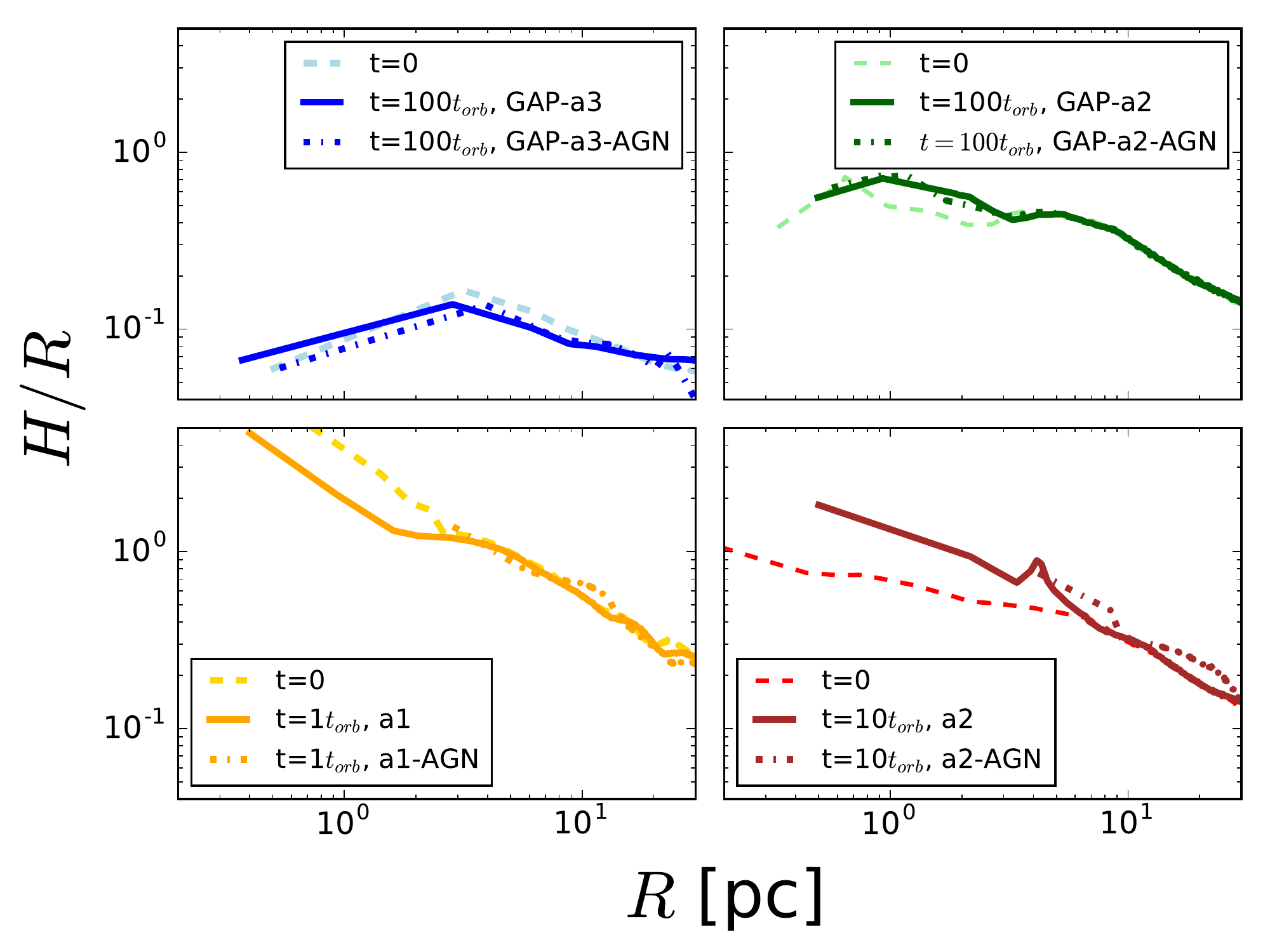}
  \caption{Thickness of the disc as a function of radius.
    Each panel corresponds to simulations with the same initial conditions.
    The top panels correspond to simulations with tidal cavity
    and the bottom panels to simulations without tidal cavity.
    In each panel $H/R$ is shown at the
    beginning of the simulation (dashed line), at the end
    of the simulation without AGN feedback (solid line)
    and at the end of the simulation with AGN feedback (dotted line).
    For systems with and without tidal cavity AGN
    feedback does not have a strong effect on the thickness of the disc.
  }
  \label{HR}
\end{figure*}

The thickness of the disc remains almost unchanged by AGN
feedback for simulations with and without a tidal cavity (Fig.~\ref{HR}).

\begin{figure*}
  \includegraphics[width=340pt]{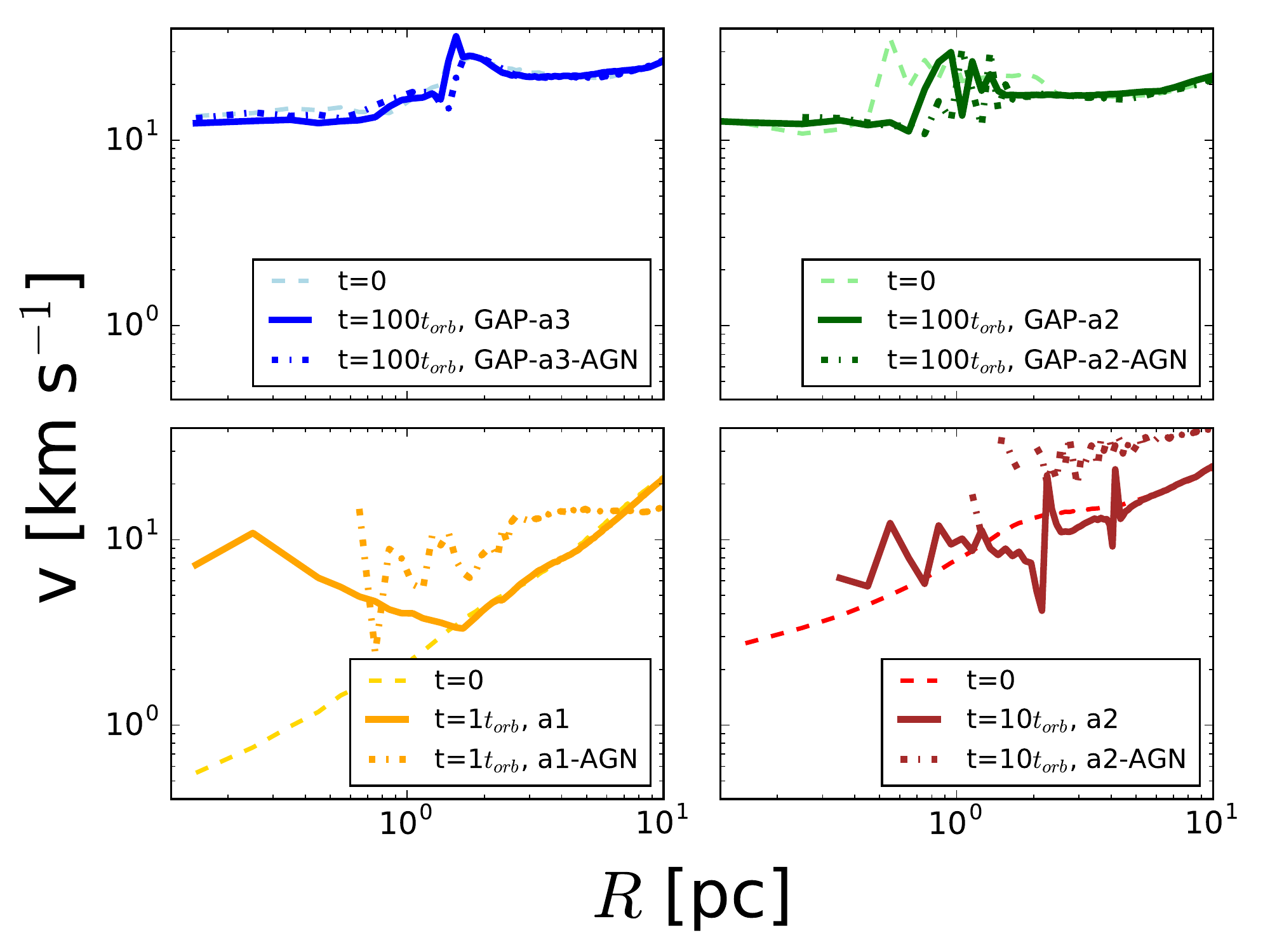}
  \caption{Rotational velocity of the disc as a function of radius.
    Each panel corresponds to simulations with the same initial conditions.
        The top panels correspond to simulations with tidal cavity
    and the bottom panels to simulations without tidal cavity.
    In each panel $v$ is shown at the
    beginning of the simulation (dashed line), at the end
    of the simulation without AGN feedback (solid line)
    and at the end of the simulation with AGN feedback (dotted line).
    In systems with tidal cavity (top panels) AGN
    feedback does not change significantly the rotational velocity
    of the gas, however, in systems without tidal cavity (bottom panels)
    AGN feedback generates strong winds that collide with the disc and
    modify the rotational velocity.
  }
  \label{Vgas}
\end{figure*}

  However, the rotational velocity of the disc
  can vary by a factor three in systems without tidal cavity
  due to AGN feedback (bottom panels Fig.~\ref{Vgas}).
  When this feedback is active the velocity of the gas
  close to the binary is not determined solely by the gravitational
  potential in that region but also by the winds generated by the
  feedback. In contrast, in simulations with tidal cavity (where
  AGN feedback does not generate strong winds that collide against the disc)
  the rotational velocity of the gas is about the same, regardless of the presence of AGN feedback
   (upper panels Fig.~\ref{Vgas}).
  
\begin{figure*}
  \includegraphics[width=350pt]{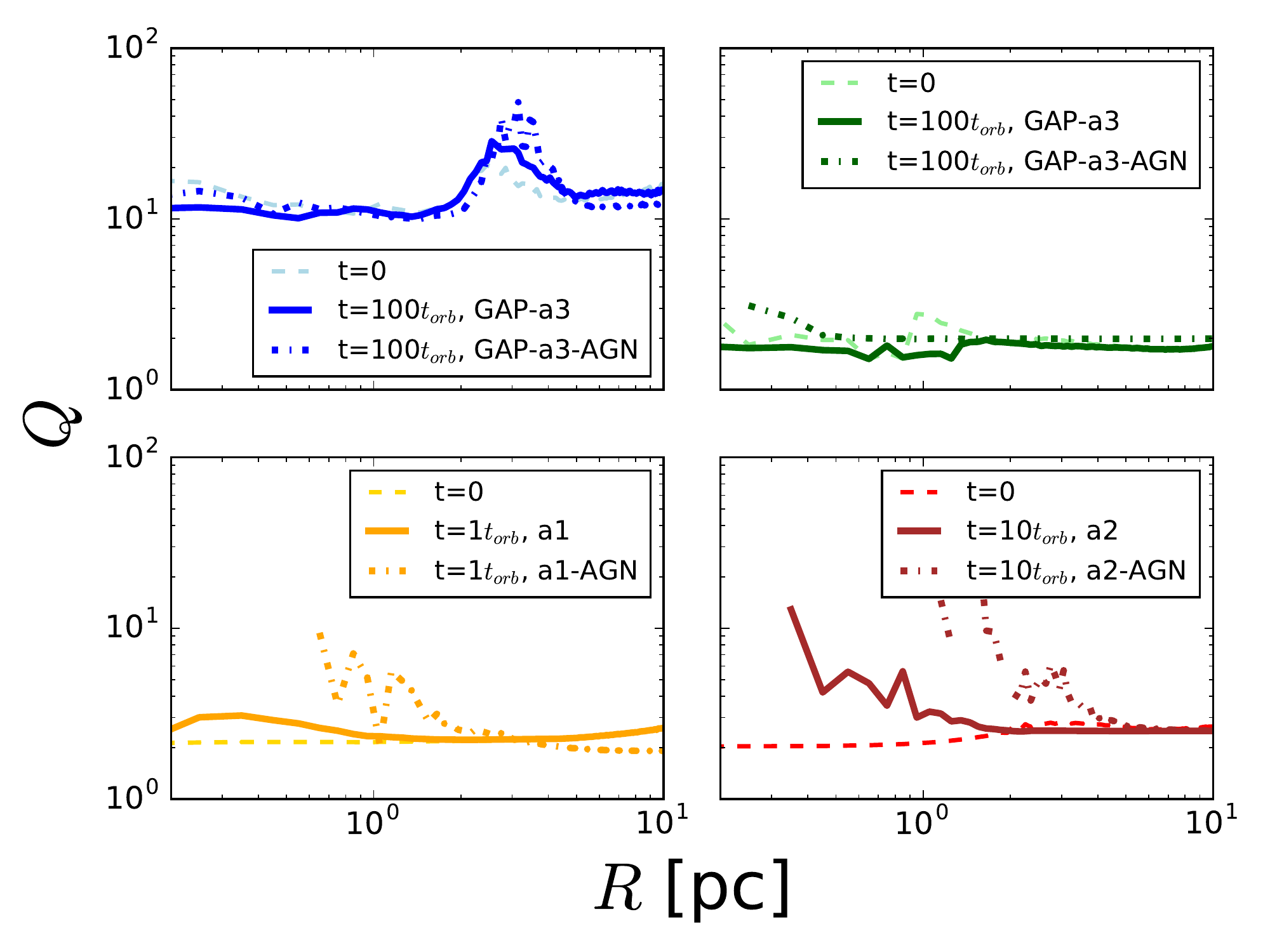}
  \caption{Stability parameter of Toomre $Q$ as a function of radius.
    Each panel corresponds to simulations with the same initial conditions.
    The top panels correspond to simulations with tidal cavity
    and the bottom panels to simulations without tidal cavity.
    In each panel $Q$ is shown at the
    beginning of the simulation (dashed line), at the end
    of the simulation without AGN feedback (solid line)
    and at the end of the simulation with AGN feedback (dotted line).
    For simulations with an initial tidal cavity (top panels)
    AGN feedback does not change significantly the stability of
    the disc. For simulations without an initial tidal cavity (bottom
    panels) AGN feedback increases the value of $Q$ in the inner
    region of the disc by a factor of $\sim10$.
  }
  \label{QtR}
\end{figure*}

 Regarding the stability of the disc, we find that AGN feedback
  can increase the value of $Q$. This increment is larger
  in systems without initial cavity, up to
  one order of magnitude (upper panels in Fig.~\ref{QtR}).
  This means that AGN feedback does not push the disc toward instability. 
  Therefore, in our setup we do not expect star formation to occur within the
  circumbinary disc.

\subsection{Outflows and Inflows}

The interaction between the AGN wind
and the circumbinary disc is different for systems with and without tidal
cavity. This results in different kinematics of the
outflows/inflows produced by AGN feedback.

\begin{figure}
  \includegraphics[width=\columnwidth]{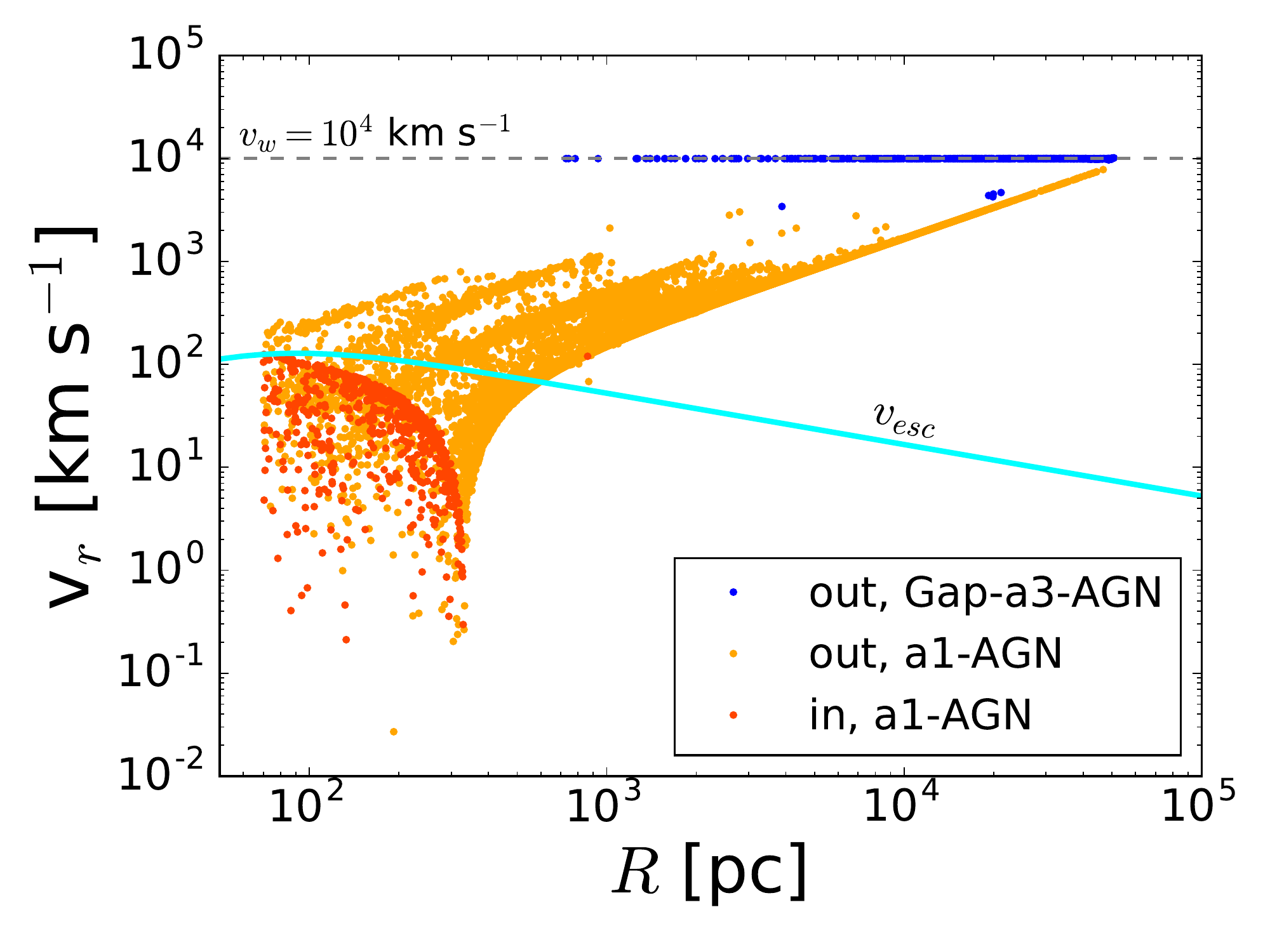}
  \caption{Radial velocity of the outflows/inflows at different radii
    after $\sim$2 Myr.
    Yellow (orange) points correspond to the absolute value of the radial velocity
    of the outflows (inflows) in simulation {\color{orange} a1-AGN}.
    Blue points correspond to the radial velocity of the outflows in simulation
    {\color{darkblue} Gap-a3-AGN}.
    The velocity of the AGN wind $v_{\rm w}$ is represented by a grey dashed line
    and the escape velocity $v_{\rm esc}$ by a cyan curve.
    The outflows generated by AGN feedback
    in the simulation with tidal cavity ({\color{darkblue} Gap-a3-AGN}) can freely escape
    without interacting with the circumbinary disc, therefore the outflows remain
    with their initial velocity $v_{\rm w}=10^4$ km s$^{-1}$. Instead, in the simulation
    without tidal cavity ({\color{orange} a1-AGN}) AGN wind crashes against
    the circumbinary disc, generating outflows/inflows with a wider distribution of velocities.
    Also, a portion of the outflows generated in simulation {\color{orange}a1-AGN}
    have velocities smaller than $v_{\rm esc}$ meaning that this gas will return
    as inflows toward the circumbinary disc.
    }
  \label{VR_R}
\end{figure}

In Fig.~\ref{VR_R} we show the absolute value of
the radial velocity of particles outside the circumbinary
disc against their distance from the center of mass of the binary.
In the same figure we plot the escape velocity of the gas
$v_{\rm esc}=(2GM(R)\,/\,R)^{1/2}$ with $M(R)$ the total mass
(Bulge + Disc + Binary) enclosed by a sphere of radius $R$.
The AGN wind in simulation
{\color{darkblue} GAP-a3-AGN} can freely escape from the disc
maintaining its initial velocity $v_{\rm w}=10^4$ km s$^{-1}$.
In contrast, in simulation {\color{orange} a1-AGN} the interaction
of the AGN wind with the circumbinary disc results in a more complex
kinematics of the outflow/inflow. Also, in simulation {\color{orange} a1-AGN},
a fraction of the outflow  does not reach
a velocity greater than the escape velocity of the system ($v_{\rm esc}$)
meaning that this gas is trapped by the gravitational potential of the
stellar bulge and will inevitably return to the circumbinary disc.

Even though the environment around the circumbinary disc
is not realistically modelled (because there is no extra gas and the presence
of stars is modelled by a static spherical potential) the difference in
the kinematics of the outflows/inflows for the two binary-disc regimes
suggests that observations of the broad and narrow absorption lines of AGN
could contain information on how the winds ejected from the accretion
discs of a SMBHB interact with the circumbinary disc.
However, any relevant conclusion about this link will need more numerical
studies that take into account the effects of gas and stars around the
circumbinary.

\section{CONCLUSIONS}
In this work we have studied, using numerical
simulations, the effect of AGN feedback on
the shrinking rate of SMBHBs embedded in
gaseous circumbinary discs.
For this we have implemented in the SPH code Gadget-3 a
model for AGN feedback, following \citet{Choi2012},
which includes: a flux prescription for accretion
on to the SMBHs, the formation of a single-velocity wind
with a mass load computed from the accretion rate
on to the SMBHs, pressure from electron scattering
and photo-ionization and the effect of photo-ionization and Compton
cooling/heating on the internal energy of the gas.

Using this implementation we have explored the effect
of the AGN feedback in two regimes of the SMBHB shrinking rate:
1) one where the binary opens a low density cavity in the disc
(tidal cavity) and therefore experiences a slow shrinking
and 2) another where the binary is not able to open a
tidal cavity and therefore shrinks rapidly.

The main results of this study are:
\begin{itemize}
\item As expected, for simulations without
  AGN feedback, when a tidal cavity
  forms the binary separation remains
  the same after $2\times10^2$ orbits and
  for binary-disc systems where no
  tidal cavity is formed the binary shrinks
  from parsec to milli-parsec scale in 1-10 orbits.  
\item When AGN feedback is active
  in a binary-disc system with a tidal cavity,
  the evolution of the binary separation and the
  structure of the disc are not affected by
  the presence of the AGN feedback because 
  the wind launched from the SMBHB can escape
  perpendicularly to the circumbinary disc through
  the tidal cavity without crashing against the disc.
\item When AGN feedback is active in a system
  without tidal cavity the SMBHs accrete at high rates. The winds are strong and 
   collide against the disc, pushing the gas away from the binary. This opens a
  ``feedback cavity'' which results in the stalling of the binary migration.
\item The radial velocity of the outflows produced
  by the winds in the regime with tidal cavity is the wind
  initial velocity $v_{\rm w}$. In the regime without
  tidal cavity, as the winds interact with the circumbinary disc,
  the radial velocity of the outflows cover a wider range and, in the
  presence of an external potential, a fraction of the displaced material
  returns to the disc, creating inflows as well as outflows.
\end{itemize}

  From these results we conclude that if a SMBHB is embedded
  in an environment where its dynamical evolution is dominated
  by gas rather than stars, and if the SMBHB is not expected
  to open a tidal cavity in the gaseous environment, then AGN feedback
  is the most important factor determining the orbital evolution of
  a SMBHB. This is because it pushes gas away from the binary opening a
  ``feedback cavity'', thus forcing the binary to migrate slower.
  AGN feedback can therefore stop the migration of the SMBHB.
  However, if the SMBHB opens a tidal cavity, then AGN feedback
  does not play a crucial role on the orbital evolution of the SMBHB.

  In this context, we may expect SMBHBs to be part of a population of
  double peaked lines AGNs, double compact cores \citep[e.g.][]{Rodriguez2006}
  or quasi periodic quasars \citep[e.g.][]{Graham2015} that will not
  enter on the regime of gravitational wave emission until they become
  part of a new galaxy merger, where the binary can become part of a
  triplet \citep{Bonetti2017}, or until some inflow of gas reaches the binary,
  driving the further shrinking of the binary \citep{Dotti2015,MaureiraFredes2018,Goicovic2018}.

  However, further research has to be done to assess the relevance of AGN
  feedback in the full cosmological landscape of the evolution of SMBHs.
  For example, we need to understand in more detail the
  typical environment of parsec scale SMBHBs and how the characteristic
  of this environment depends on the previous history of the galaxy merger.


\section*{Acknowledgements}

LDV and MV acknowledge  funding  from  the  European  Research
Council  under  the  European  Community's  Seventh  Framework
Programme (FP7/2007-2013 Grant Agreement no. 614199, project
``BLACK'').  This  work  was  granted  access to  the  HPC
resources of CINES under the allocation A0020406955 made
by GENCI. 



\bibliographystyle{mnras}
\bibliography{biblio}


\appendix

\section{MBH Accretion/Feedback Implementation Test}
\label{code_app}
To test the accretion and feedback implementation, we follow the evolution of a
$Q$-stable gas disc rotating around a single SMBH of
$10^6\,M_{\odot}$.

Initially the disc follows a Mestel surface density profile with
a radius of 6 pc, a scale height of 0.5 pc and a total mass comparable
with the mass of the SMBH. We use this SMBH + disc simulation to determine
if feedback can regulate taccretion and to test the results against
physical expectations.

The parameters that we vary in the test simulations are: accretion radius
$R_{\rm acc}$, total number of gas particles sampling the disc $N_{\rm gas}$
and the number of neighbours of the SPH scheme $n_{\rm nbg}$.

\subsection{Flux Accretion Radius}
\label{AccretionApp}
\begin{figure}
	\includegraphics[width=\columnwidth]{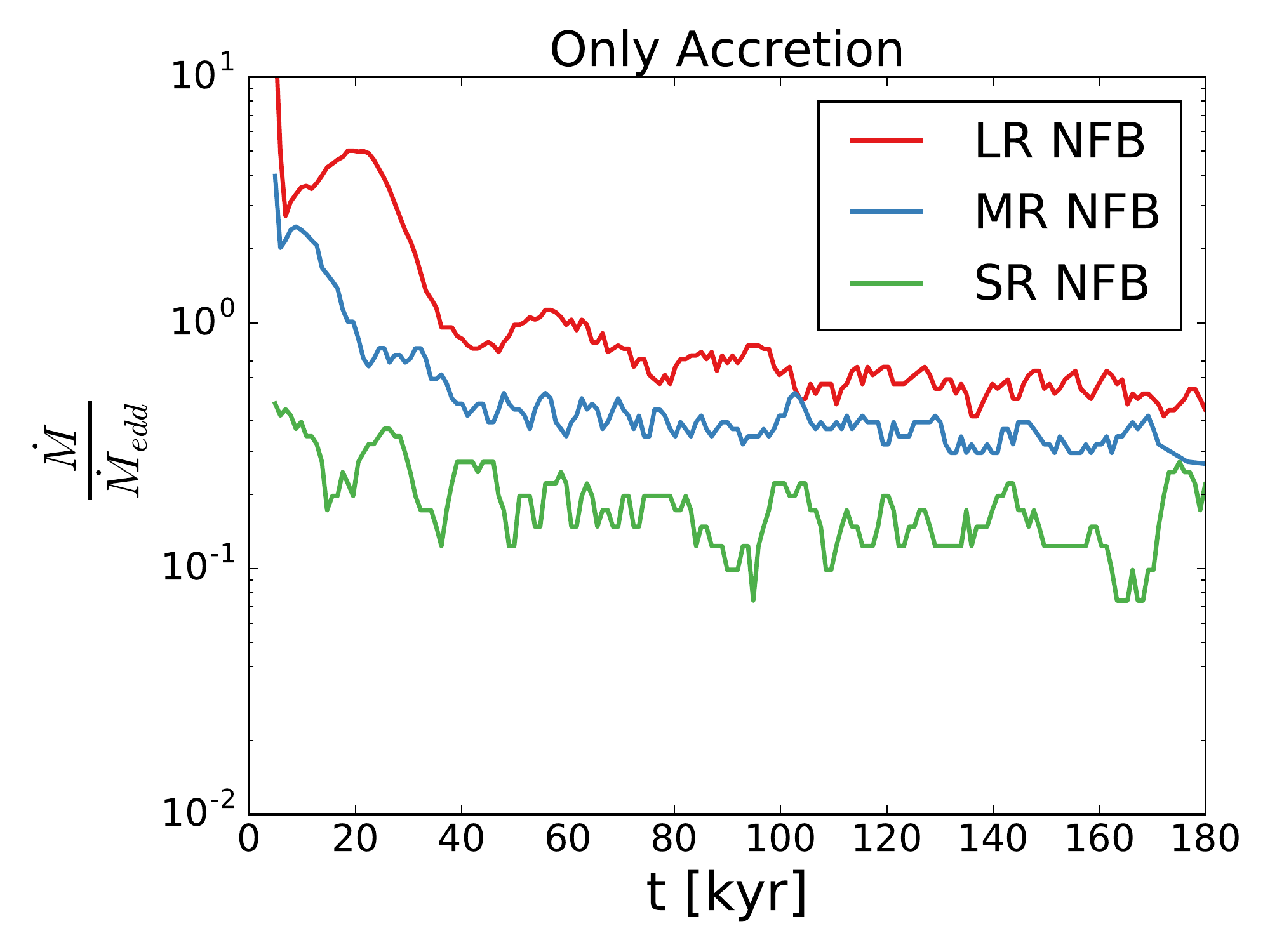}
        \caption{Accretion rate normalized by Eddington accretion rate
          against time in 3 simulations with different accretion radii.
          $R_{\rm acc} = $0.4 pc (red line or ``LR NFB''), $R_{\rm acc} = $0.1 pc
          (blue line or ``MR NFB'') and $R_{\rm acc} = $0.04 pc (green line or ``SR NFB'').
          These simulations do not include AGN feedback.
          For smaller accretion radii the
          accretion rate of the black hole decreases.}
    \label{test_Racc_NOAGN}
\end{figure}

In Fig.~\ref{test_Racc_NOAGN} we show the ratio between the accretion
rate of the SMBH and its Eddington rate for simulations without
feedback. Each colour represents a simulation with different $R_{\rm acc}$.
The values that we use for $R_{\rm acc}$ are $R_{\rm acc} = 0.4$ pc
(red line or ``LR''), $R_{\rm acc} = 0.1$ pc (blue line or ``MR'')
and $R_{\rm acc} = 0.04$ pc (green line or ``SR'').
All these accretion radii are larger than the typical size of
$\sim0.01$ pc of an accretion $\alpha$-disc \citep{SS1973}
that is stable against self-gravity and that is orbiting around
a SMBH of $10^6\,M_{\odot}$ accreting at the Eddington rate
\citep{KolSyun1980}. Different accretion radii result in different
accretion rates. This difference can be easily understood: when the
MBH has a larger region to eat from (larger $R_{\rm acc}$) it eats more.

\begin{figure}
	\includegraphics[width=\columnwidth]{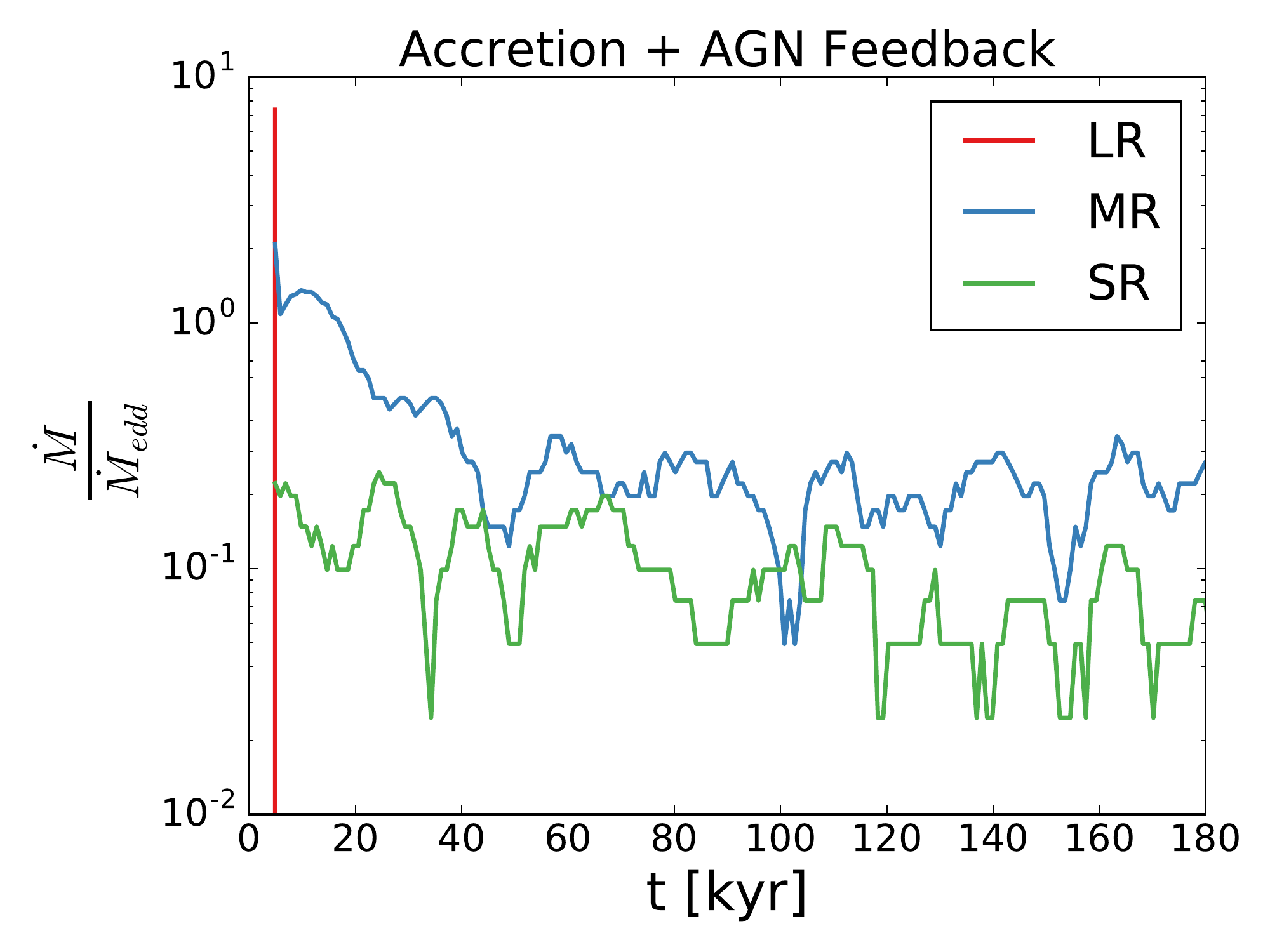}
        \caption{Accretion rate normalized by Eddington accretion rate
          against time in 3 simulations with different accretion radii.
          $R_{\rm acc} = $0.4 pc (red line or ``LR''), $R_{\rm acc} = $0.1 pc
          (blue line or ``MR'') and $R_{\rm acc}=$0.04 pc (green line or ``SR'').
          All these simulations include the AGN feedback implementation.
          In simulations with small accretion radius
          (``SM'') and medium accretion radius (``MR'') the accretion rate
          of the black hole are comparable thanks to the AGN feedback.
          Instead, for the large accretion radius (``LR''), the initial
          spike in accretion generates strong feedback that totally stops
          any further accretion on to the black hole.
          }         
    \label{test_Racc_AGN}
\end{figure}

\begin{figure}
  \includegraphics[width=78pt]{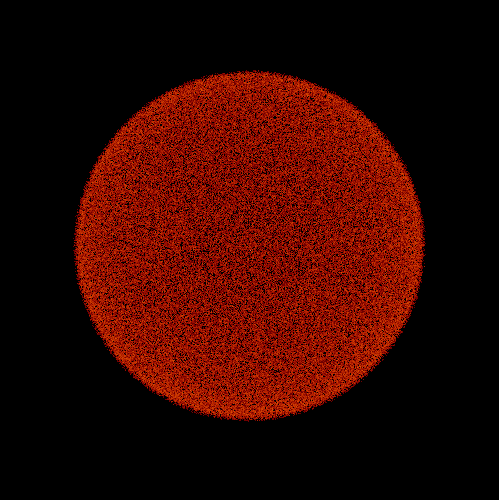}\includegraphics[width=79pt]{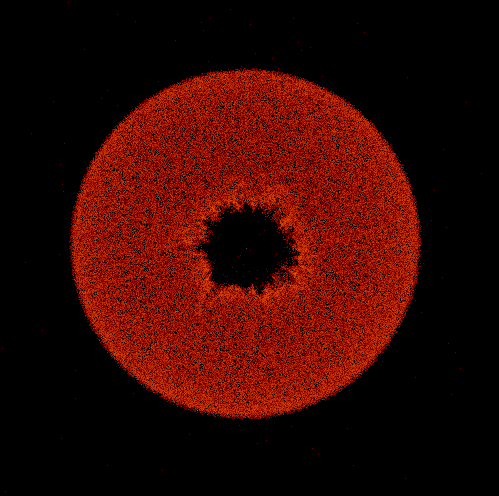}\includegraphics[width=81pt]{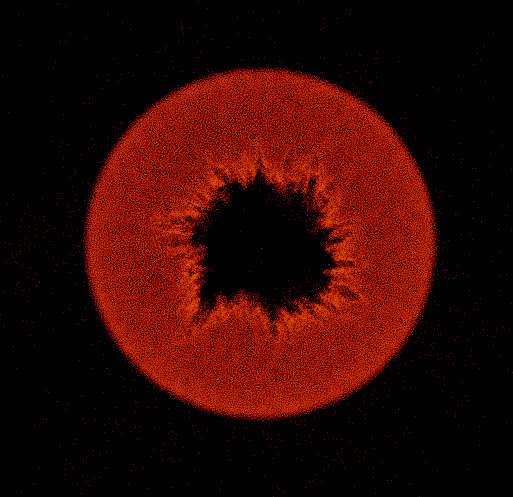}
  \caption{Density projection of simulation ``LR'' at three different times.
    The intense feedback in simulation ``LR'' pushes away all the gas around the black hole stopping any further accretion. 
  }
    \label{gas_blow_away}
\end{figure}

 When AGN feedback is active  (Fig.~\ref{test_Racc_AGN})
the mean fractional difference between simulations ``SR'' and ``MR'' is of 
$\sim 89$\% which is slightly smaller than the mean fractional difference
without AGN feedback $\sim 96$\%.
For the simulation with the largest accretion radius (``LR'') the initial AGN feedback is
so intense that the gas is blown away and the SMBH accretion stalls.
This is even clearer in Fig.~\ref{gas_blow_away}, where
the strong AGN feedback in simulation ``LR'' destroys the disc.

\begin{figure}
	\includegraphics[width=\columnwidth]{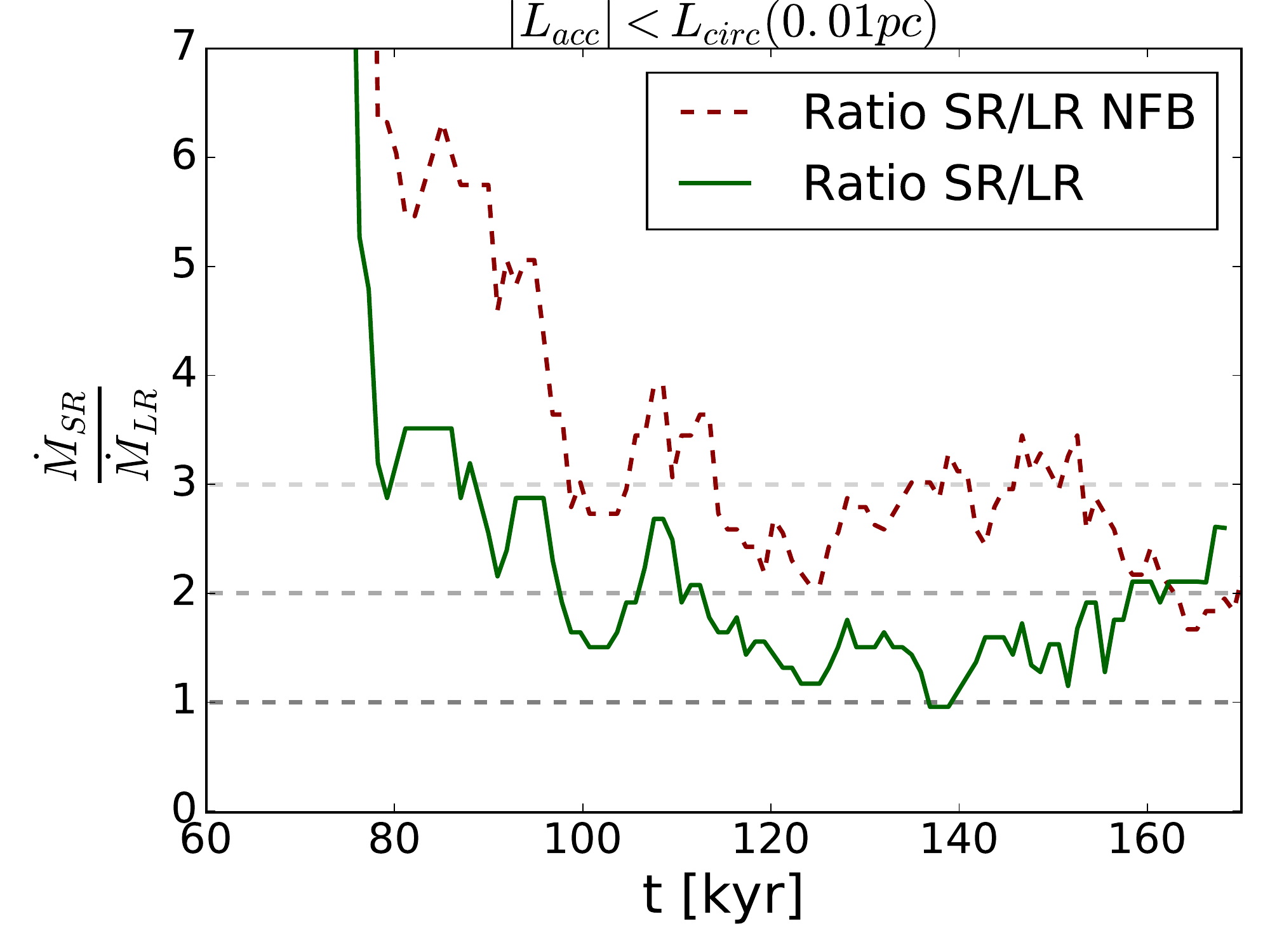}
        \caption{Ratio between accretion rate in simulations
          with a small accretion radius ``SR'' ($R_{\rm acc}=0.01$)
          and simulations with a large accretion radius ``LR''
          ($R_{\rm acc}=0.4$). The dashed dark-red curve corresponds
          to simulations with accretion but without AGN feedback and the
          solid dark-green curve corresponds to simulations with accretion and
          AGN feedback. An angular momentum restriction on to the black holes
          results in a more similar accretion rate between simulations with
          a small accretion radius (``SR'') and a large accretion radius (``LR'').
        }
    \label{ratio_rate}
\end{figure}

\begin{figure}
	\includegraphics[width=\columnwidth]{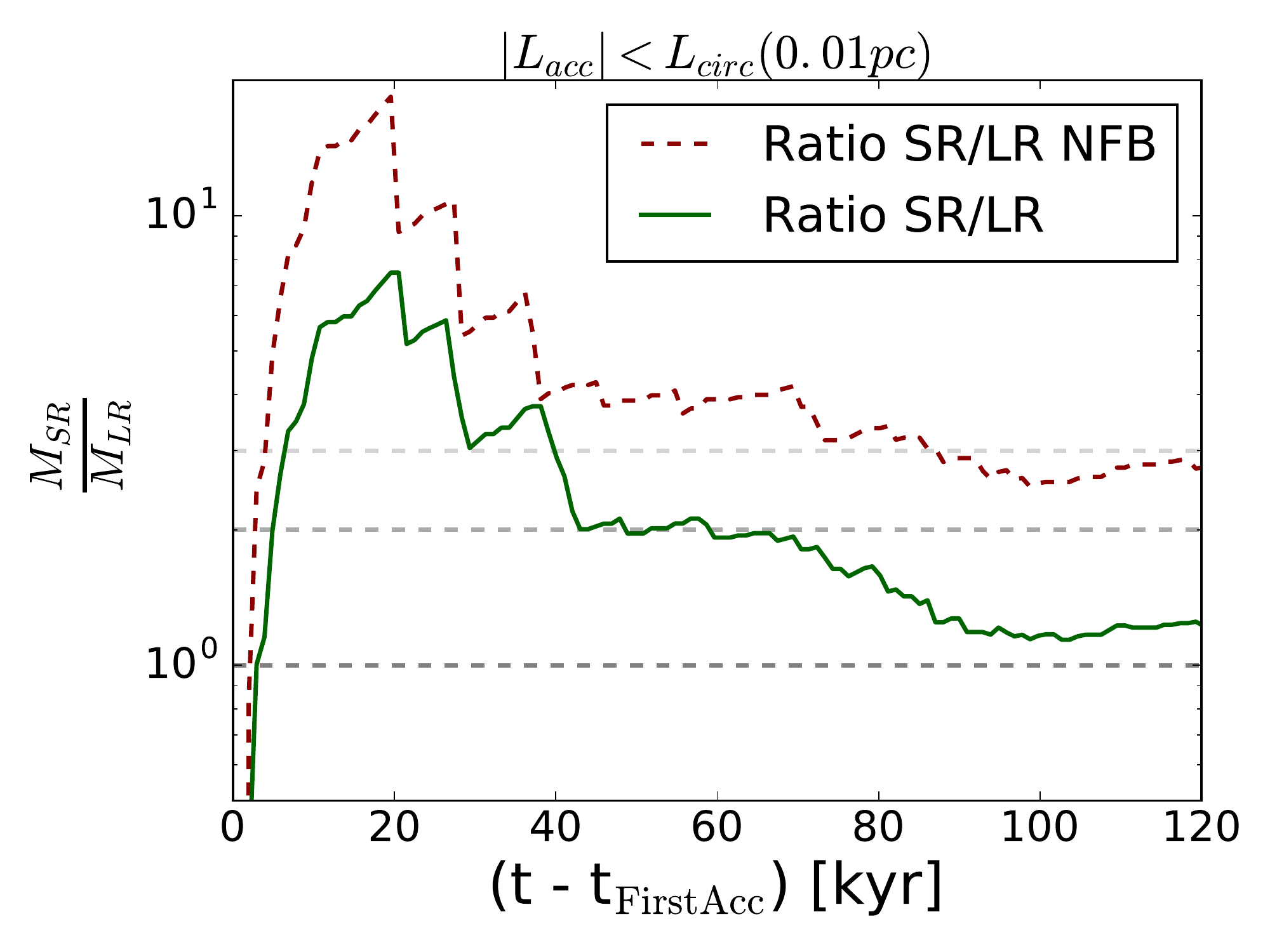}
        \caption{Ratio between the total accreted mass in simulations
          with a small accretion radius ``SR'' ($R_{\rm acc}=0.01$)
          and simulations with a large accretion radius ``LR''
          ($R_{\rm acc}=0.4$). The dashed dark-red curve corresponds
          to simulations with accretion but without AGN feedback and the
          solid dark-green curve corresponds to simulations with accretion and
          AGN feedback. The total mass accreted by
          the black hole is comparable between simulations with different a accretion
          radius when the AGN feedback is active and an angular momentum restriction
          is used for the accretion.
        }
    \label{total_mass}
\end{figure}

In order to minimize the difference on the accretion rate due to difference
in the size of the accretion radius, we include another condition on accretion:
we only allow the SMBH to accrete gas that has an angular momentum ($L_{\rm acc}$)
smaller than the angular momentum of a circular orbit of radius 0.01 pc around
the SMBH. This radius corresponds to the maximum radius of an $\alpha$-disc
stable against self gravity \citep{KolSyun1980}. The extra condition allows
the SMBH with larger accretion radius (``LR'') to continue its accretion without
blowing away all the gas (Fig.~\ref{ratio_rate}). In fact, its accretion results in a total
accreted mass comparable with the accreted mass in simulation ``SR'', where $R_{\rm acc}$
is ten times smaller (Fig.~\ref{total_mass}).

\subsection{Mass Resolution and the Effect of Neighbour Number in Viscosity}

We study how the number of particles that are
sampling the gas disc can affect the accretion rate. For this purpose,
we run 5 simulations with a different number of particles and a different
number of neighbours used by the SPH scheme.

\begin{figure}
	\includegraphics[width=\columnwidth]{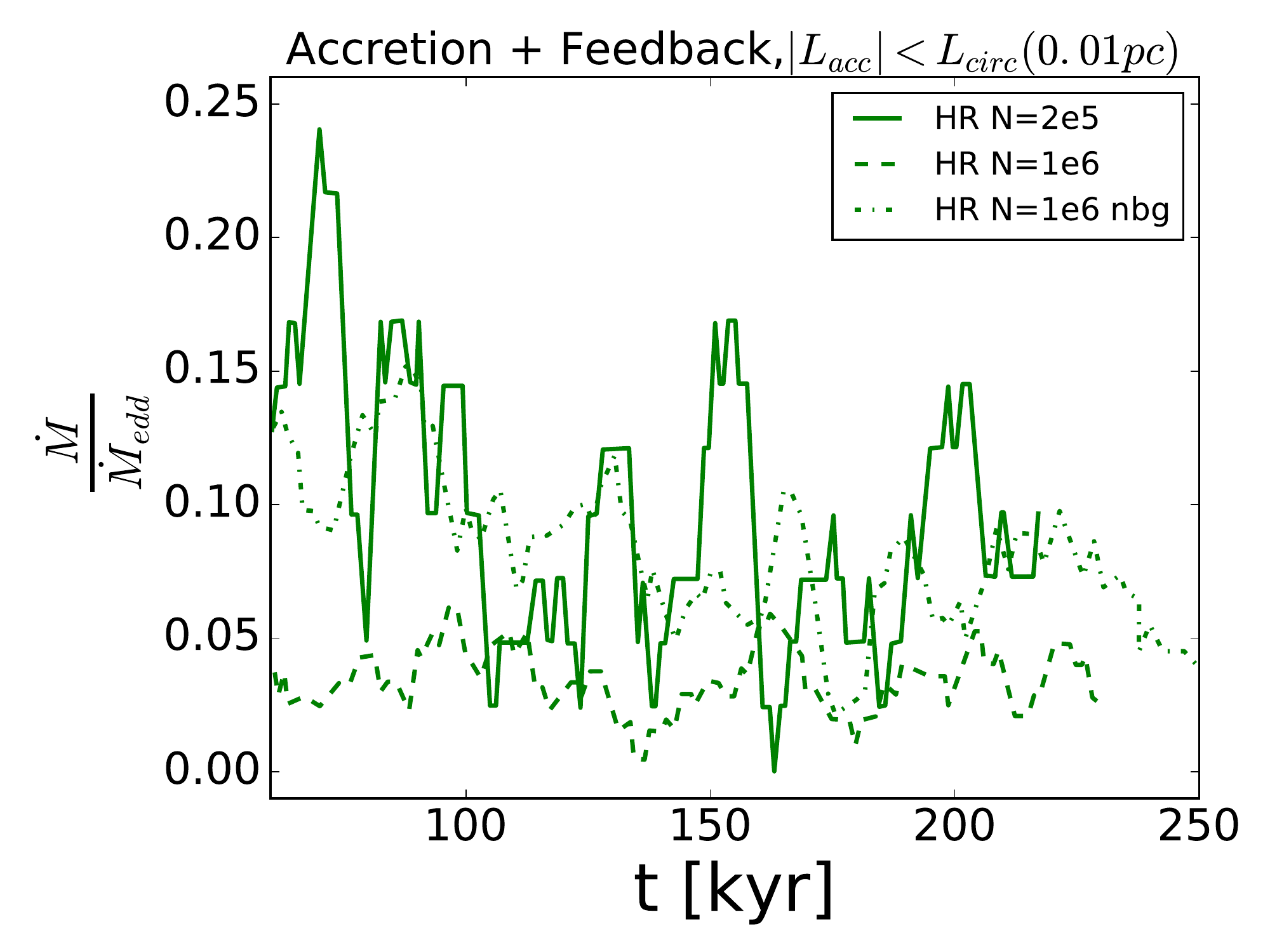}
        \caption{Accretion rate normalized by Eddington accretion rate against time. The solid line corresponds
          to a simulation with $2\times 10^5$ gas particles, the dashed line to a simulation with $10^6$ gas
          particles and dotted-dashed line to a simulation with $10^6$ particles but where the number
          of neighbour of the SPH scheme has been increased in order to maintain the average smoothing
          length roughly constant.
          An increase in the number of particles, which decreases the
          average smoothing length and subsequently the viscosity, results in a decrease in the
          accretion rate (dashed line) while if we also increase the number of neighbours,
          to maintain the average smoothing length constant, the accretion rate does not change
          (dotted-dashed line). 
          }
    \label{test_number}
\end{figure}

In Fig.~\ref{test_number} we show the accretion rate of
the SMBH (normalized to the Eddington rate) in three different simulations,
all of them with the same accretion radius. The solid line corresponds to
a simulation with  $2\times10^5$ particles sampling
the gas disc and where the SPH kernel used 64 neighbours, the dashed line
corresponds to a simulation where the disc is sampled with $10^6$ particles also
with 64 neighbours and the dotted green line corresponds to
a simulation with $10^6$ particles but with 320 neighbours.
The higher resolution simulations ($N_{\rm gas}=10^6$) have a
smoother accretion rate curve. If we increase the number of particles without
changing the number of neighbours (solid line compared to dashed line)
the accretion rate decreases by a factor of $\sim$2.
This happens because  accretion on to the SMBH is controlled by the viscosity
in the disc, which can be estimated as
$\nu_{\rm SPH} = 0.1\,\alpha_{\rm AV}\,\left< h_{\rm SPH} \right>\,c_s$ where
$\alpha_{\rm AV}$ is a constant value that is an input parameter of Gadget-3 to
model the viscosity, $\left< h_{\rm SPH} \right>$ is the mean smoothing length
and $c_s$ is the sound speed of the disc \citep{LodatoPrice2010}. Therefore, for a
higher number of particles, and constant number of neighbours, the smoothing
length will decrease resulting in a decrease of the viscosity and, as a
consequence, of the accretion rate.
Accordingly, in a simulation with a larger number of particles sampling
the disc ($N_{\rm gas}=10^6$) imposing a larger number of neighbours in the
SPH scheme allows to recover an accretion rate similar to the one obtained 
with fewer gas particles ($N_{\rm gas}=2\times10^5$).


\bsp	
\label{lastpage}
\end{document}